# Development of Hydrogen Bonding Magnetic Reaction-based Gene Regulation through Cyclic Electromagnetic DNA Simulation in Double-Stranded DNA

**Suk Keun Lee[1]*, Dae Gwan Lee[2], Yeon Sook Kim[3]**

[1] Department of Oral Pathology, College of Dentistry, Gangneung-Wonju National University, Gangneung; and Institute of Hydrogen Bonding Magnetic Reaction Gene Regulation (Private research center, not registrated), Daejeon, South Korea

[2] Department of Mathematics and Big Data Science, Kumoh National Institute of Technology, Gumi, South Korea

[3] Department of Dental Hygiene, College of Health and Medical Sciences, Cheongju University, Cheongju, South Korea

**Short title:** Cyclic electromagnetic DNA simulation effect on dsDNA

*Corresponding Authors:

Suk Keun Lee, DDS, MSD, PhD.

Department of Oral Pathology, College of Dentistry, Gangneung-Wonju National University, 7, Jukheon-gil, Gangneung, 25457 South Korea.

E-mail : sukkeunlee2@naver.com

---

# Development of Hydrogen Bonding Magnetic Reaction-based Gene Regulation through Cyclic Electromagnetic DNA Simulation in Double-Stranded DNA

**Abstract**

The proton-magnetic reaction is commonly used in MRI machines with a strong magnetic field of over 1 T, while this study hypothesized that the electron magnetic reaction of hydrogen could affect the hydrogen bonds of double-stranded DNA (dsDNA) at a low magnetic field below 0.01 T. The electron magnetic reaction in hydrogen bonds is referred to as hydrogen bonding magnetic reaction (HBMR) in this study. The goal is to develop a HBMR-based gene regulation (HBMR-GR) system, which can safely modulate the conformations and functions of dsDNAs[1]. The polarities of DNA base pairs are derived from the relative electrostatic charge between purines and pyrimidines, which become positively and negatively charged, respectively. The Pyu dsDNAs with pyrimidine(s)-purine(s) sequences, ds3T3A, ds3C3G, and ds3C3A, showed stronger DNA hybridization potential, increased infrared absorption at 3400-3200 $cm^{-1}$, and a unique DNA conformation in HPLC analysis compared to the corresponding Puy dsDNAs. To target the three-dimensional structure of dsDNA based on the DNA base pair polarities, one can use cyclic electromagnetic DNA simulation (CEDS) with approximately 25% efficiency for randomly oriented dsDNAs. This study found that the most effective magnetic exposure times for inducing an electric potential in A-T and G-C base pairs were 280 and 480 msec, respectively. CEDS was found to induce sequence-specific hybridization of target oligo-dsDNAs in 0.005M NaCl solution and sequence-specific conformation of oligo-dsDNAs in 0.1M NaCl solution. It was found that the Pyu oligo-dsDNAs were more responsible for the hybridization and conformational changes by CEDS than the Puy oligo-dsDNAs. CEDS decreased ethidium bromide (EtBr) DNA intercalation and spermidine DNA condensation depending on CEDS time in the binding assay. The results also included that the Pyu oligo-dsDNAs were more responsible for CEDS by forming stable and unique conformation of oligo-dsDNA than the Puy oligo-dsDNAs. CEDS using a corresponding restriction site or promoter sequence enhanced *in vitro* restriction endonuclease (RE) digestion and *in vitro* RNA transcription, respectively, with the pBluescript II vector. Furthermore, the reporter protein assay, which uses β-galactosidase or GFP plasmid vectors to detect protein production from E. coli, was also improved by CEDS using a corresponding promoter sequence in a sequence-specific manner. Therefore, it is postulated that the low-level HBMR-based CEDS can enhance the hybridization potential of oligo-dsDNAs and subsequently lead to the unique DNA conformation required for the initiation of various DNA functions.

**A concept of DNA base pair polarity**

The DNA double helix consists of four nucleotides: two purines (deoxyadenosine triphosphate and deoxyguanosine triphosphate) and two pyrimidines (deoxythymidine triphosphate and deoxycytidine triphosphate). The sequence consisting of these nucleotides determines genetic signals and is supported by hydrogen bonds between two complementary DNA strands. The structure of double-stranded DNA (dsDNA) is characterized by the consistent arrangement of purine and pyrimidine bases, sugar molecules, and phosphate groups. This arrangement creates a unique signaling circuit that is stimulated and buffered by hydrogen bonds, resulting in the control of the electrical charge in its purine/pyrimidine bases and phosphodiester groups.

In DNA base pairs, purines possess a greater number of hydrogen donors and contain a cationic imidazole ring in comparison to pyrimidines. It is hypothesized that the purine is positively charged by transferring the electron from the pyrimidine moiety to the imidazole moiety, when the π-resonance-facilitating electron is charged by the hydrogen bonding associated with the complementary pyrimidine. Consequently, purines, such as adenine (A) and guanine (G), can assume a positive charge, while pyrimidines, such as thymine (T) and cytosine (C), can assume a negative charge in the context of hydrogen bonding within DNA base pairs.

In a DNA strand, electrons in the pyrimidine-dependent phosphate group move with greater ease to the purine-dependent phosphate group, where electrons are partially depleted due to the associated cationic purine. In contrast, electrons in the purine-dependent phosphate group resist moving down to the pyrimidine-dependent phosphate group, where electrons are saturated by the associated anionic pyrimidine. It appears that there is a boundary between purine and pyrimidine in a DNA strand, which can regulate the electrical circuit in DNA sequences. It is therefore proposed that the entire dsDNA sequences be divided into Pyu (pyrimidine(s)-purine(s)) dsDNA segments, distinguished by graphical symbols of the DNA bases (Fig. 1A).

This concept allows clear visualization of potential electrostatic interactions between adjacent base pairs, as well as the boundary between pyrimidine(s) and purine(s) in forward and reverse DNA strands. The flow of current from pyrimidine(s) to purine(s) is enhanced by the forward attraction of the purine-dependent phosphate group, while the flow of current from purine(s) to pyrimidine(s) is hindered by the reverse attraction of purine-dependent phosphate group of a DNA axis. Therefore, we hypothetically assume the existence of an electric current boundary between purine(s) and pyrimidine(s) in a DNA strand and draw a boundary line mark between purine(s) and pyrimidine(s) in the symbolic graph of dsDNA (Fig.1 B).

We tentatively assigned $+2\alpha$, $-3\alpha$, $+3\alpha$, and $-2\alpha$ base pair polarities to A, C, G, and T, respectively, based on the number of hydrogen bonds in their pairs (where $\alpha$ is a constant value per hydrogen bond) (see Fig. 1 B). Therefore, the electric charge of pyrimidines and purines in a Pyu dsDNA segment can be calculated and combined into $\alpha$ value in the DNA base pair polarity program (Korean patent: KR101287040B1).

This study demonstrates that the whole dsDNA can be divided into Pyu dsDNA segments, which are the basic units of the genetic code, to preserve the potential energy of DNA base pairs and to induce optimal DNA conformation for the native DNA roles of RNA and DNA transcription (Fig. 1 C). And in the same line, this study aims to identify the biological significance of Pyu dsDNA segments compared to Puy dsDNA segments.

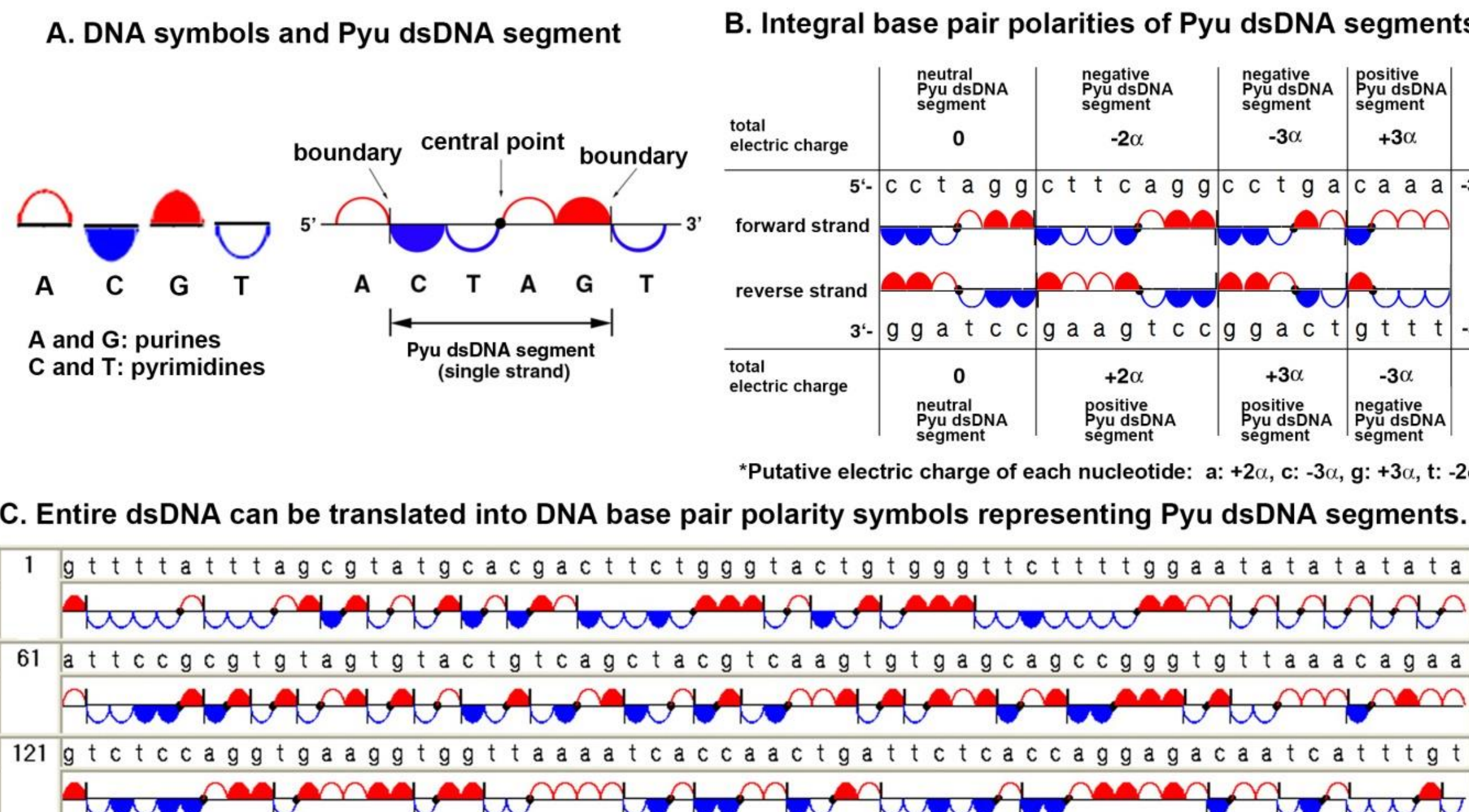

**Figure 1.** Symbolic illustration of DNA base pair polarities in dsDNA. A: Symbols are used for each DNA base, A, C, G, and T, along with the design of a representative Pyu dsDNA segment. B: Pyu ssDNA strands pair complementary, resulting in an antagonistic total electric charge for a Pyu dsDNA. C. The entire dsDNA codes can be divided into Pyu dsDNA segments, illustrated by the symbols of DNA base pair polarities.

**Different properties between Pyu and Puy dsDNA segments**

***1) Hybridization difference between Pyu and Puy oligo-dsDNAs***

To determine the difference in hybridization status between Pyu and Puy oligo-dsDNAs of different sizes, the oligo-ssDNAs, ssTnAn and ssAnTn (n=3, 4, 5, 6, 7, 8, 10, and 12) were purchased from Cosmogenetech Co. Ltd. (Korea, Seoul) and dissolved in 0.1 M NaCl solution at a concentration of 100 μmol/mL. The samples were heated to 90℃ for 10 min and gradually cooled to room temperature before analysis by the EtBr intercalation electrophoresis method.

The pre-electrophoresis EtBr stain did not allow visualization of small dsDNAs shorter than ds8A8T and ds8T8A due to weak EtBr retention. However, post-electrophoresis EtBr staining produced a blurred image showing the small dsDNA. These results suggest that Pyu dsDNAs, especially dsTnAn, underwent monomeric hybridization up to ds8T8A and dimeric hybridization in 10T10A and 12T12A. In contrast, Puy dsDNAs, specifically dsAnTn, showed monomeric hybridization up to ds5A5T and tended to form dimeric hybridization from ds6A6T to ds12A12T. On the other hand, the dimeric hybridization products of Pyu dsDNAs, specifically dsT10A10 and dsT12A12, showed a higher density of EtBr staining in both pre-electrophoresis and post-electrophoresis EtBr staining compared to those of Puy dsDNAs (Fig. 2).

The data suggest that Pyu dsTnAn (n= 3, 4, 5, 6, 7, 8, 10, and 12) have stronger monomeric and dimeric hybridization than Pyu dsAnTn, indicating the higher DNA hybridization potential of Pyu oligo-dsDNAs compared to Puy oligo-dsDNAs.

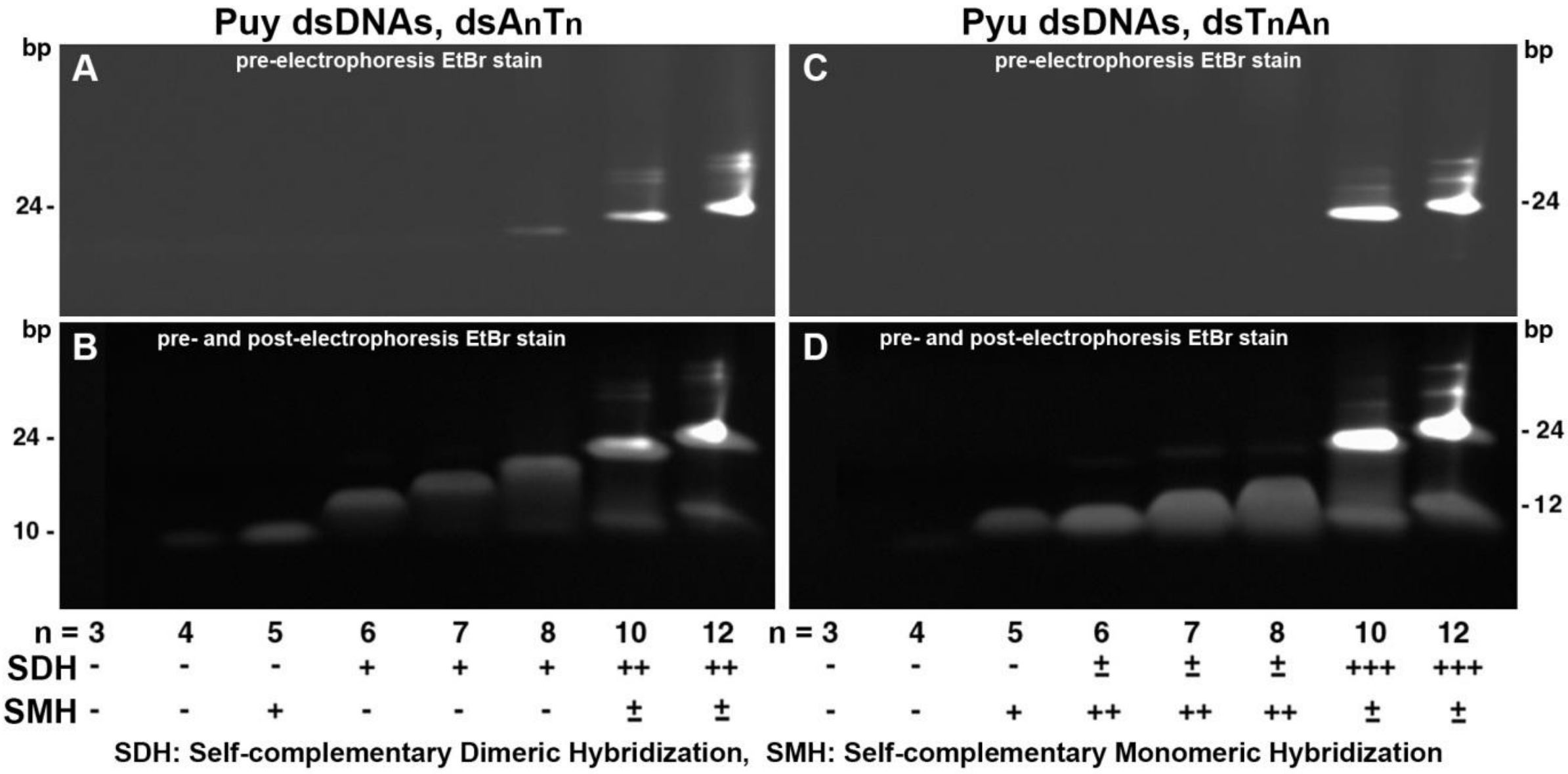

**Figure 2.** Hybridization of oligo-dsAnTn and oligo-dsTnAn (n= 3, 4, 5, 6, 7, 8, 10, and 12) was detected using EtBr stain. A and C: Pre-electrophoresis EtBr stain, B and D: Pre- and post-electrophoresis EtBr stain. SDH refers to self-complementary dimeric hybridization, while SMH stands for self-complementary monomeric hybridization. *This figure represents three or more repeated experiments.

Since the oligo-ssDNA pairs such as ssCnAn and ssTnGn (n=7, 8, 10 and 12), ssAnCn and ssGnTn (n=7, 8, 10 and 12) may exhibit dimeric hybridization in the absence of self-complementary monomeric hybridization, they were used in the hybridization assay to compare the dimeric hybridization strength between Pyu and Puy oligo-dsDNAs, dsCnAn and dsAnCn (n=7, 8, 10 and 12). Each oligo-ssDNA pair was purchased and dissolved in 0.1 M NaCl solution at 100 μmol/mL, heated to 90℃ for 10 minutes and gradually cooled to room temperature to obtain dsCnAn and dsAnCn (n=7, 8, 10, and 12) with complete dimeric hybridization. The samples were then analyzed by the EtBr intercalation electrophoresis method.

As the value of n increased, dsCnAn (n=7, 8, 10, and 12) showed significantly higher efficiency in EtBr intercalation compared to dsAnCn (n=7, 8, 10, and 12). Specifically, ds7C7A showed 90% more EtBr intercalation than ds7A7C, and ds8C8A showed 68.9% more EtBr intercalation than ds8A8C. Similarly, ds10C10A showed 25% more EtBr intercalation than ds10A10C, and ds12C12A showed 6% more EtBr intercalation than ds12A12C (Fig. 3). The results indicate that Pyu oligo-dsDNAs, dsCnAn (n=7, 8, 10, and 12) showed stronger hybridization and recruited more EtBr intercalation than Puy oligo-dsDNAs, dsAnCn (n=7, 8, 10, and 12).

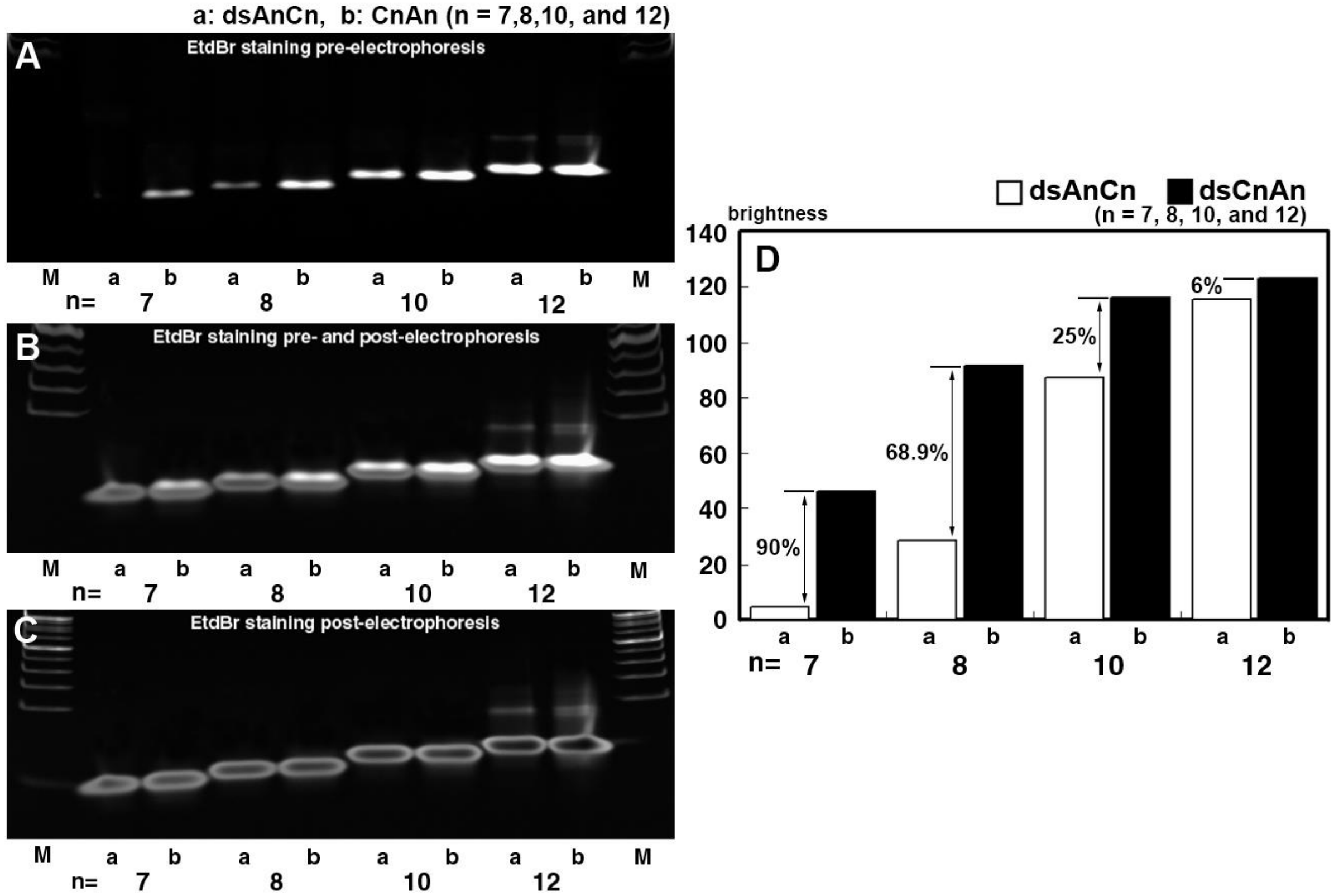

**Figure 3.** The EtBr intercalation was used to detect hybridization of oligo-dsAnCn and oligo-dsCnAn (n = 7, 8, 10, 12). Pre-electrophoresis EtBr staining is shown in panel A, while panels B and C show both pre- and post-electrophoresis EtBr staining, respectively. Panel D shows a comparison graph of EtBr intercalation between the dsAnCn and dsCnAn groups from panel A. This figure is representative of replicate experiments performed over three times.

### *2) Conformational differences between Pyu and Puy oligo-dsDNAs*

The conformational differences between the Pyu and Puy oligo-dsDNAs could be determined by their distinctive UV260 nm absorption in high performance liquid chromatography (HPLC) analysis. The different oligo-dsDNA pairs were dissolved in a 0.1 M NaCl solution at 100 pmole/μL separately and then hybridized by heating to 95°C and slowly cooling to room temperature. The oligo-dsDNA sample (50 μL) was analyzed by HPLC using a reverse phase column packed with silica beads (Daisogel SP-300-5P, Osaka Soda Co., Ltd., Japan) and a running buffer of 0.1 M NaCl solution at a flow rate of 0.3 mL/min and a temperature of 30°C.

For simple six sequence-long oligo-dsDNAs consisting of six A-T or G-C pairings, the Pyu dsDNAs (ds3T3A and ds3C3G) showed a larger peak area by 4.8% and 29.7%, respectively, compared to the Puy dsDNAs (ds3A3T and ds3G3C) (Fig. 4 A and D). In addition, the Pyu dsDNAs composed of six pairs of A-T, ds3T3A and ds3(TA), exhibited a larger peak area by 144.2% and 72.8%, respectively, than the Pyu dsDNAs composed of six pairs of G-C, ds3C3G and ds3(CG) (Fig. 4 B and E). In addition, the Pyu dsDNAs consisting of three dsDNA segments, ds3(TA) and ds3(CG), showed peak areas 70.7% and 141.2% larger, respectively, than the Pyu dsDNAs consisting of one dsDNA segment, ds3T3A and ds3C3G (Fig. 4 C and F).

Each simple oligo-dsDNA shows a unique HPLC peak pattern depending on the number and sequence of A-T and G-C pairs. Among the six oligo-dsDNAs, each containing the same number of A-T and G-C pairs, the Pyu dsDNAs exhibited larger peak areas than the Puy dsDNAs. In addition, the Pyu dsDNAs composed of A-T pairs had larger peak areas than those composed of G-C pairs, and the Pyu dsDNAs composed of three dsDNA segments had larger peak areas than those composed of one dsDNA segment.

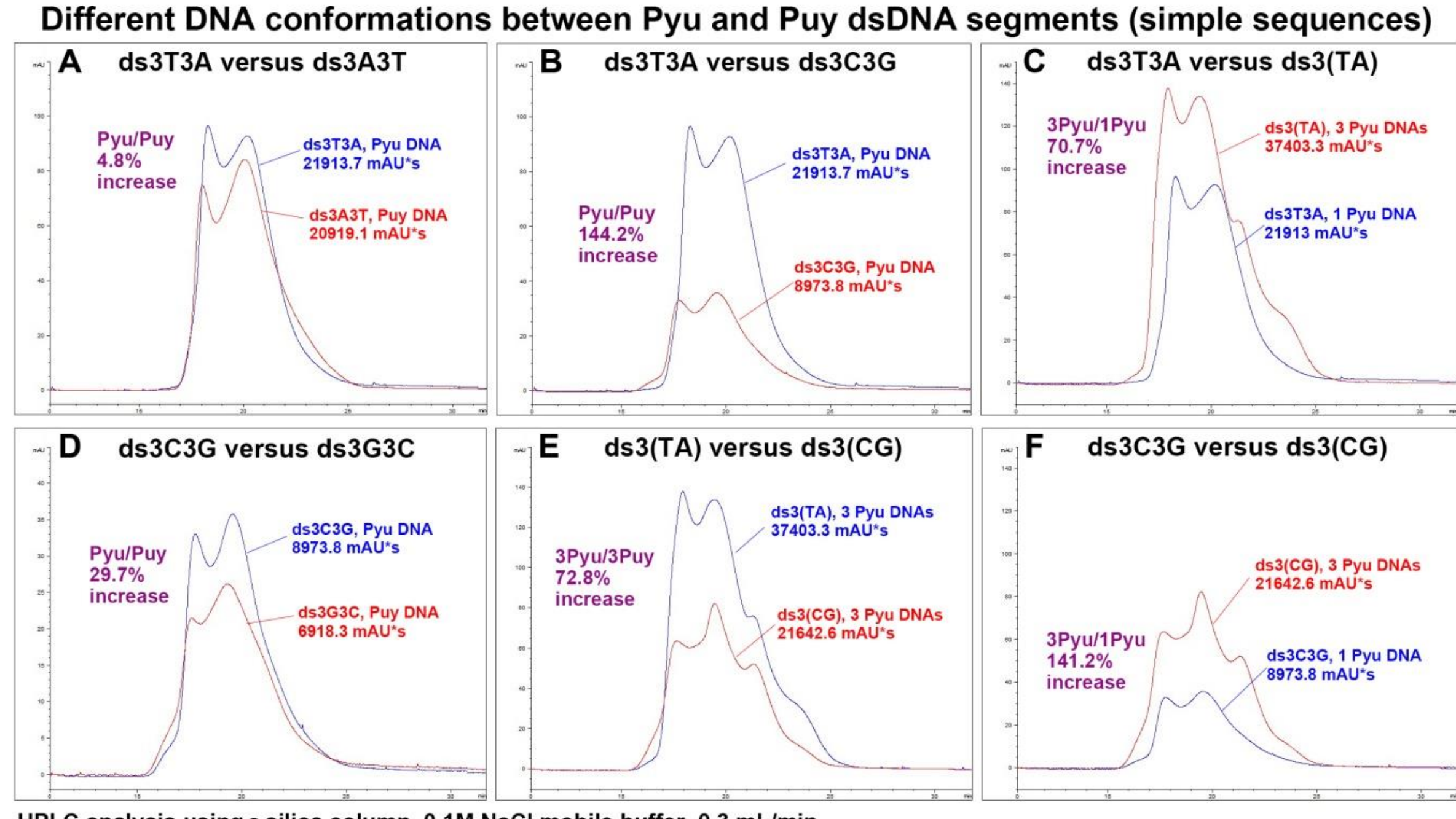

**Figure. 4.** HPLC analysis for six sequence-long simple oligo-dsDNAs, including ds3T3A, ds3A3T, ds3C3G, ds3G3C, ds3(TA), and ds3(CG) in 0.1M NaCl solution. Noted the contrasting DNA conformation between the Pyu and Puy oligo-dsDNA, showing a trend that the single Puy dsDNA segments, ds3A3T and ds3G3C, showed smaller conformational size than the single Pyu dsDNA segments, ds3T3A and ds3C3G, while the multiple Pyu dsDNA segments, ds3(TA) and ds3(CG), showed larger conformational size than the single dsDNA segments, ds3T3A and ds3C3G.

When comparing complex six sequences-long oligo-dsDNAs including ds3C3A and ds3A3C, it was observed that ds3C3A had a faster retention time on the reverse phase column but a smaller peak area than ds3A3C by 17.2% (Fig. 5 A). But the ds3C3A had a similar retention time but a 39.7% smaller peak area than ds3C3G (Fig. 5 D). Based on these results, it can be concluded that the complex Pyu dsDNA, ds3C3A is a more condensed form compared to the complex Puy dsDNA, ds3A3C and the simple Pyu dsDNA, ds3C3G.

When comparing complex six sequence-long oligo-dsDNAs, the Pyu dsDNA, dsCTCGAG (XhoI restriction site) showed a 12.6% larger peak area than the Puy dsDNA, dsGAGCTC (SacI restriction site) (Fig. 5B). The Pyu dsDNA, dsTCTAGA (XbaI restriction site) showed 9.1% larger peak area than the Puy dsDNA, dsAGATCT (BgIII restriction site) (Fig. 5C). The Pyu dsDNA, dsCTTAAG (AfIII restriction site) showed 147.1% larger peak area than the Puy dsDNA, dsGAATTC (EcoRI restriction site) (Fig. 5E). The Pyu dsDNA, dsTTCGAA (BstBI restriction site) showed a slightly faster retention time and 2.7% increase in peak area compared to the Puy dsDNA, dsAAGCTT (HindIII restriction site) (Fig. 5 F). The peak areas of the six sequence-long palindromic dsDNAs with different endonuclease restriction sites varied depending on the characteristics of the Pyu and Puy dsDNAs, although each set of dsDNAs had the same number of A-T and G-C pairs.

When analyzing DNA conformation using HPLC peak area (mAU*s), a larger peak area may indicate that the DNA structure has expanded and is absorbing more UV260 nm. Conversely, a smaller peak area may indicate that the DNA structure has contracted and absorbs less UV260 nm. The above Pyu oligo-dsDNAs have a larger peak area than the comparable Puy oligo-dsDNAs, indicating that they have a more extended DNA structure due to the stronger hydrogen bonding and weaker base stacking in the Pyu oligo-dsDNAs than in the Puy oligo-dsDNAs. The different conformation of Pyu and Puy dsDNAs may represent the different energy state of each other, and then this was simply identified by the contrasting DNA base pair polarity between the Pyu and Puy dsDNAs in this study.

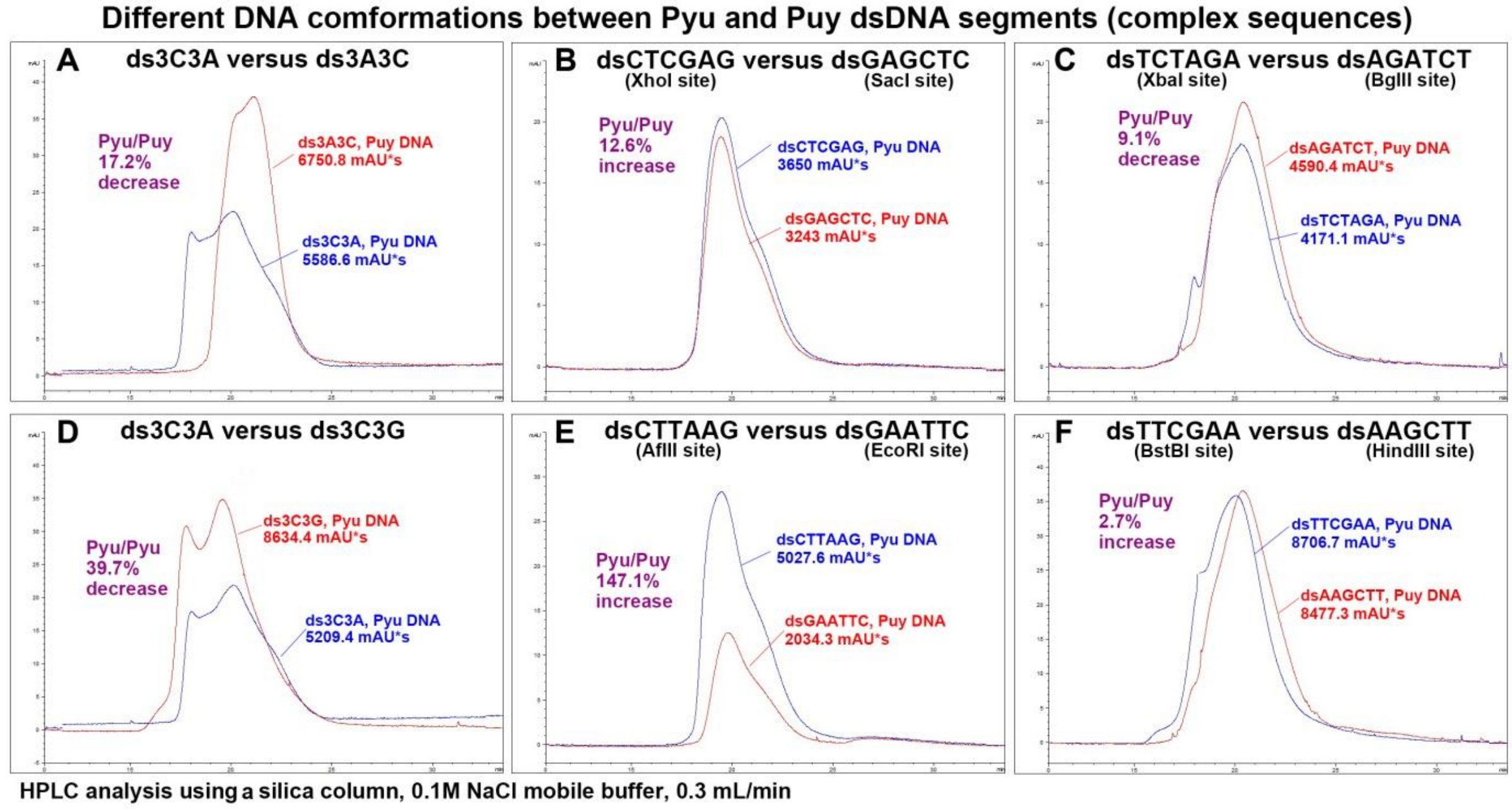

**Figure 5.** HPLC analysis for comparing complex Pyu and Puy oligo-dsDNA pairs, including ds3C3A versus ds3A3C (A), dsCTCGAG versus dsGAGCTC (B), dsTCTAGA versus dsAGATCT (C), ds3C3A versus ds3C3G (D), dsCTTAAG versus dsGAATTC (E), and dsTTCGAA versus dsAAGCTT (F) in 0.1M NaCl solution. The conformational differences between the complex Pyu and Puy oligo-dsDNAs were irregular and contrasting, but tended to be sequence-specific.

### *3) FT-IR difference between Pyu and Puy oligo-dsDNAs*

The magnetically treated water (MTW) showed the increase of infrared (IR) absorption as the increase of hydrogen bonding strength[2]. Similarly, to compare the hydrogen bonding strength between the Pyu and Puy oligo-dsDNAs, the following oligo-dsDNA sets were prepared in 0. 1 M NaCl solution as previously described and analyzed by FT-IR (UATR Two, Perkin Elmer, USA): ds3T3A and ds3A3T, ds3C3G and ds3G3C, ds3C3A and ds3A3C, dsCCCGGG and dsTTAGGG, dsCTCGAG and dsGAGCTC, and dsTCTAGA and dsAGATCT.

In FT-IR analysis, the IR absorption of NH-O hydrogen bonds in oligo-dsDNA was compared to OH-O hydrogen bonds in double distilled water (DDW), 0.1M NaCl, and 1M NaCl solutions. However, the Pyu dsDNAs (ds3T3A, ds3C3G, dsCTCGAG, and dsTCTAGA) showed higher IR absorption at 3400-3200 $cm^{-1}$ than the Puy dsDNAs (ds3A3T, 3G3C, dsGAGCTC, and AGATCT), respectively (Fig. 6 A, B, E, and F), but the Pyu dsDNA, ds3C3A showed lower IR absorption than the Pyu dsDNA, ds3A3C (Fig. 6 C). On the other hand, the telomere repeat sequence consisting of a Pyu dsDNA, dsTTAGGG, showed lower IR absorption than the similar Pyu dsDNA, dsCCCGGG (Fig. 6 D).

The data show that oligo-dsDNAs with a Pyu dsDNA segment have higher IR absorption than those with a Puy dsDNA segment, indicating that the former has stronger hydrogen bonding in DNA hybridization than the latter. However, ds3A3C, which has a larger DNA conformation in HPLC analysis than ds3C3A, showed higher IR absorption than ds3C3A. This suggests that both IR absorption and DNA conformation of oligo-dsDNA are variable in a sequence-specific manner.

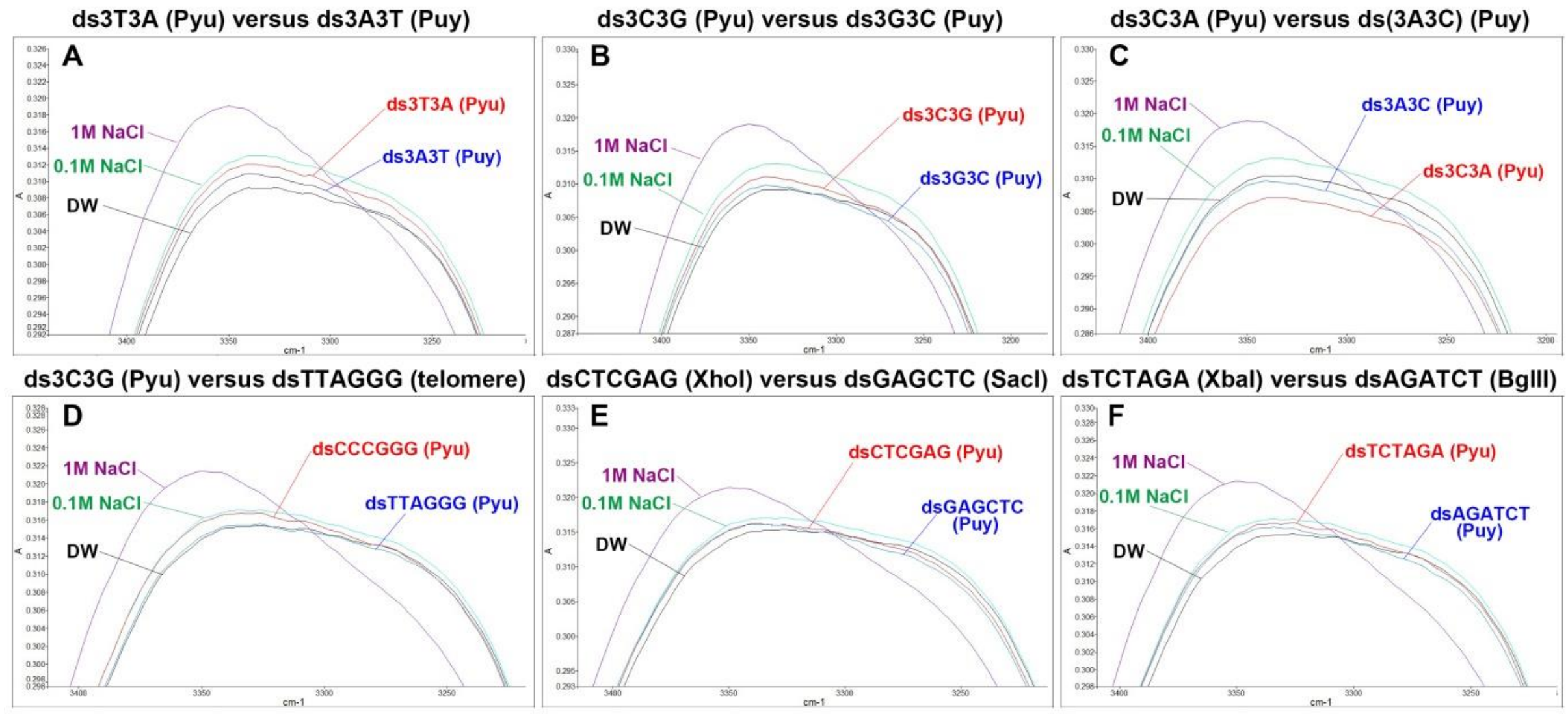

**Figure 6.** FT-IR analysis was performed on different sets of Pyu and Puy oligo-dsDNAs at 3400-3200 $cm^{-1}$. The Pyu oligo-dsDNAs, including ds3T3A, ds3C3G, dsCTCGAG, and dsTCTAGA, showed higher IR absorption than the corresponding Pyu oligo-dsDNAs, including ds3A3T, ds3G3C, dsGAGCTC, and dsAGACTC, while the Pyu oligo-dsDNA, ds3A3C showed higher IR absorption than the Pyu oligo-dsDNA, ds3C3A. Furthermore, the telomere repeat sequence composed of a Pyu dsDNA segment, dsTTAGGG exhibited lower IR absorption than a Pyu dsDNA segment, ds3C3G.

### Hydrogen bonding magnetic reaction (HBMR) in hydrogen bonds between DNA base pairs

Nuclear magnetic reactions can be influenced by both strong magnetic fields above 1 Tesla (T) and weaker magnetic fields, such as the geomagnetic field (about 50 μT). The external magnetic field ($B_{ext}$) that is much weaker than the internal magnetic field ($B_{int}$) in a hydrogen atom can induce a magnetic reaction in the electron, leading to the splitting of the spectral lines as a Zeeman Effect, which is also known to induce electron transfer in the associated hydrogen bonds[3-5]. These weaker fields can induce spin-orbit interaction of electrons in the hydrogen atom of water molecules, resulted in the magnetically treated water (MTW) exhibiting the increase of T2 relaxation time in NMR spectroscopy[6].

In this study, the magnetic reaction in the hydrogen bonding is referred to as HBMR (hydrogen bonding magnetic reaction). The intermolecular HBMR in the hydrogen bonds of dsDNA can be occurred by a weak external magnetic field ($B_{ext} \ll B_{int}$), and can induce the electron transfer from the hydrogen acceptor to the hydrogen donor. The putative energy (E) of HBMR in a hydrogen bond can be calculated using the Zeeman equation in a weak magnetic field:

$$\boldsymbol{E = g_l \mu_B B m_l}$$

where $g_l$ is the orbital gyromagnetic ratio, $\mu_B$ is the Bohr magneton, $B$ is the magnetic field strength, and $m_l$ is the magnetic quantum number ($m_l = 1$)[7-9]. For example, 100 Gauss of magnetic field can induce about 1.2 μeV in a hydrogen atom by Zeeman Effect, which is readily transferred into the associated hydrogen bonds depending on the direction of magnetic field, if the quantum radiation were negligible.

In this aspect, HBMR can increase the electric charge of hydrogen bonds, and the repeated HBMR is required to increase the electric potential in hydrogen bonds. It is supposed that about 200 repetitions of HBMR using 100 Gauss of magnetic field can increase the spin-orbit momentum, up to about 240 μeV in a hydrogen bond, which may be effective on the conformational changes of the associated molecules.

The hydrogen donor of hydrogen bonds in dsDNA are all N-H groups, which are more labile and show higher chemical shift in nuclear magnetic resonance spectroscopy than O-H groups present in water. Therefore, it is assumed that the hydrogen bonds of dsDNA will be responsible to magnetic field, leading to the changes of DNA conformation and functions.

**Design of decagonal or dodecagonal CEDS to target dsDNA**

To induce the HBMR effect on a dsDNA, we applied the Cyclic Electromagnetic DNA Simulation (CEDS) to cyclically target the three-dimensional structures of DNA base pairs depending on their base pair polarities and magnetic exposure time for each A-T and G-T pair. CEDS system utilizes electromagnetic bars arranged in a circular pattern to target 10 and 12 base pairs per revolution for the B- and Z-type dsDNAs, respectively. This is because the active B- and Z-type dsDNAs have approximately 10 and 12 base pairs per revolution, respectively, and then five or six pairs of electromagnetic bars are aligned in a line facing each other by focusing their magnetic field on the center of the longitudinal axis of the dsDNA. The B-type dsDNA is targeted with five pairs of electromagnetic bars (Fig. 7A), while the Z-type dsDNA is targeted with six pairs of electromagnetic bars (Fig. 7B). Thereby, CEDS is capable of targeting both B- and Z-type dsDNAs located in the center of the electromagnetic bar alignment.

CEDS uses a magnetic field of 10-30 Gauss, running from pyrimidine to purine direction for each base pair, and sequentially rotating in the counter-clockwise or clockwise direction to simulate the structure of the dsDNA for 15-30 min (see Fig. 7). The direction of the magnetic field polarity between base pairs and the circular direction of the magnetic field are not obligatory due to the chiral structures of dsDNA in three-dimensional space, but should be done regularly in relation to neighboring base pairs in dsDNA sequences, as they could be reversed in the chaotic state of randomly distributed dsDNAs.

For example, to target both strands of B-type dsACGTA and Z-type dsTTAGGG, we used the decagonal (Fig. 7A) and dodecagonal (Fig. 7B) CEDS in sequential procedures from ① to ⑩ and from ① to ⑫, respectively. Since the DNA base pairs in one or two revolutions are almost on the same plane perpendicular to the longitudinal axis of dsDNA, CEDS may be effective in targeting 6-24 base pairs of dsDNA, even in a supercoiled structure.

A-T and G-C base pairs are exposed with different magnetic exposure time, 280 and 480 msec, respectively, according to the determination of the effective exposure time for HBMR in DNA base pairs in the following experiment (see Fig. 9).

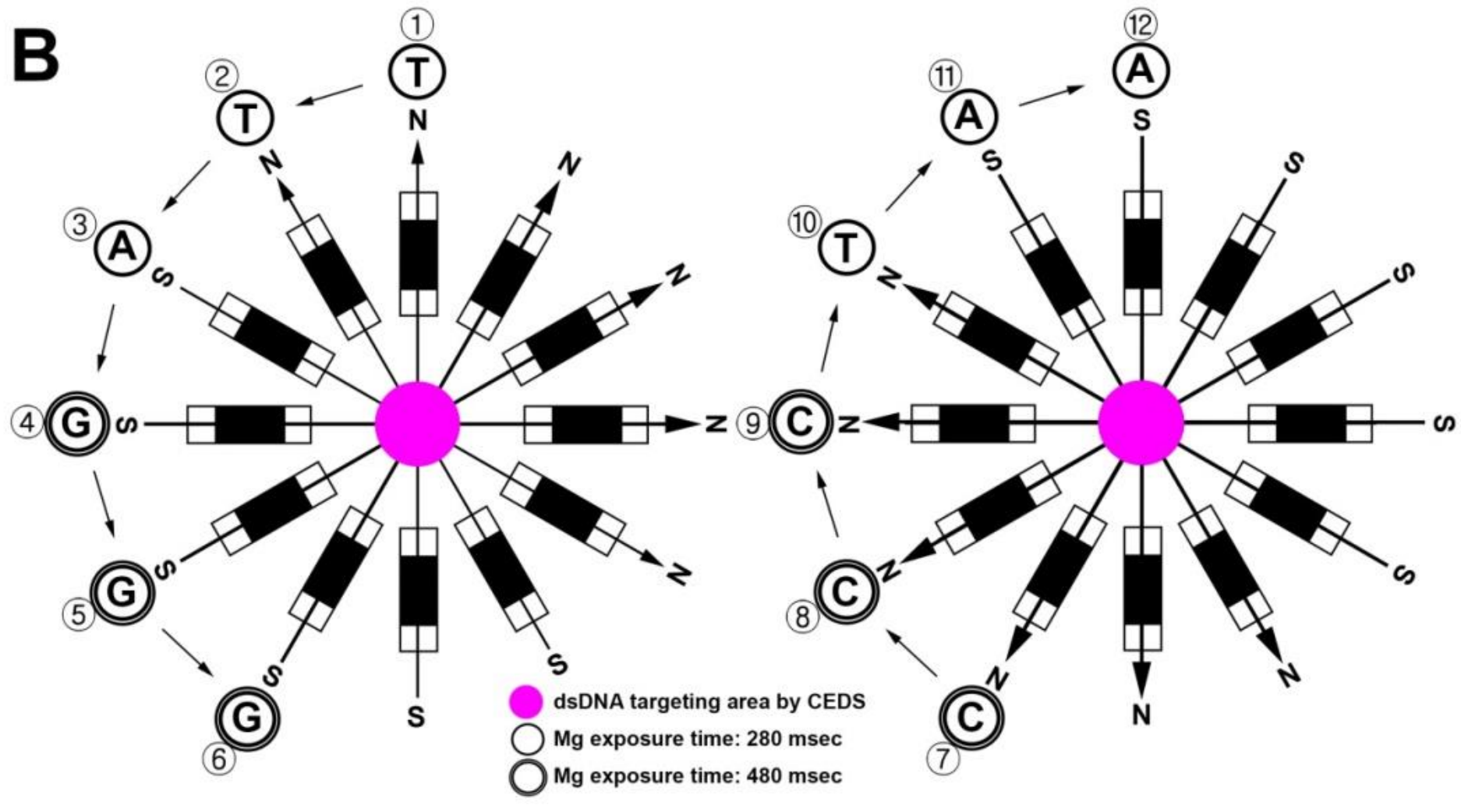

**Figure 7.** Schematic illustration of CEDS apparatus. B-type dsACGTA and Z-type dsTTAGGG (a unit of telomere) are targeted by decagonal (A) and dodecagonal (B) CEDS in the procedures of ① → ⑩ and ① → ⑫, respectively.

Two types of CEDS apparatus are available for the HBMR-based gene regulation (HBMR-GR), the first being the horizontal type and the second being the vertical type. The former has a horizontal arrangement of magnetic bars that are focused at the center of the magnetic poles, while the latter has a vertical arrangement of magnetic bars that are focused at the outward-extending magnetic field lines.

In the horizontal CEDS apparatus, CEDS reaction is available in a spherical area located at the center of the magnetic poles, and the size of the spherical area is determined by the width of the magnetic field lines generated by each pair of magnetic poles. In the vertical CEDS apparatus, CEDS reaction is available in the cylindrical area of outwardly extending magnetic field lines generated by each pair of magnetic poles, and the diameter and length of the cylindrical area are determined by the width and strength of the magnetic field lines, respectively (Korean patent: WO2013012276A2, International patent: PCT/KR2012/005790).

To regulate genes, it is more effective to stimulate both strands of the dsDNA rather than just one. This is because the dsDNA is composed of complementary forward and reverse ssDNA strands held together by hydrogen bonds with different base pair polarities. In this study, CEDS was used to influence the hybridization states and conformational changes of objective dsDNAs by using the sequences of both strands. CEDS that uses both strands of target dsDNA will be referred to as "target dsDNA sequence*-CEDS". For example, CEDS that uses both strands of ds3T3A is referred to as "3T3A*-CEDS", while CEDS that uses only one strand of ds3T3A is referred to as "3T3A-CEDS".

**Estimation of the physical efficiency of CEDS on the DNA double helix**

According to the operating principle of CEDS, the dsDNA in Figure 8A which is oriented in the +z axis direction with the sense sequence (solid line) pointing upwards and the antisense sequence (dashed line) pointing downwards, is assumed to be fully affected by the CEDS. However, such exact alignment occurs very rarely because dsDNAs have random orientations in three dimensions. The CEDS is assumed to have no effect on those dsDNAs that are placed horizontally or turned upside down. Taking these cases into consideration, the physical efficiency of CEDS on a dsDNA oriented in the (θ,φ) direction can be modeled as

$$\eta(\theta,\varphi)=\begin{cases}\cos\varphi, & 0\le\varphi<\pi/2\\ 0, & \frac{\pi}{2}\le\varphi\le\pi.\end{cases}$$

In particular, the case $\varphi = 0$ corresponds to the dsDNAs that are oriented in the +z axis direction, and therefore are affected by the CEDS with full efficiency $\eta(\theta,\varphi)=1$. The cases $\varphi = \pi/2$ and $\varphi = \pi$ correspond to the dsDNAs that are placed horizontally or turned upside down, respectively, and therefore are affected by the CEDS with efficiency $\eta(\theta,\varphi)=0$, indicating no effect by the CEDS.

By denoting the orientation of a dsDNA as a unit vector in the spherical coordinate system ($0\le\theta<2\pi$, $0\le\varphi\le\pi$, $r=1$), the random orientations of dsDNAs in three dimensions is represented by the uniform distribution over the unit sphere. Therefore, the average physical efficiency of CEDS on randomly distributed dsDNAs in three dimensions is given by

$$\frac{1}{4\pi\,1^2}\int_0^{2\pi}\int_0^{\frac{\pi}{2}}\cos\varphi\cdot 1^2\sin\varphi\;d\varphi\;d\theta=\frac{1}{4}.$$

In summary, the theoretical efficiency of CEDS on randomly distributed dsDNAs in three dimensions is limited by 25% (Fig. 8).

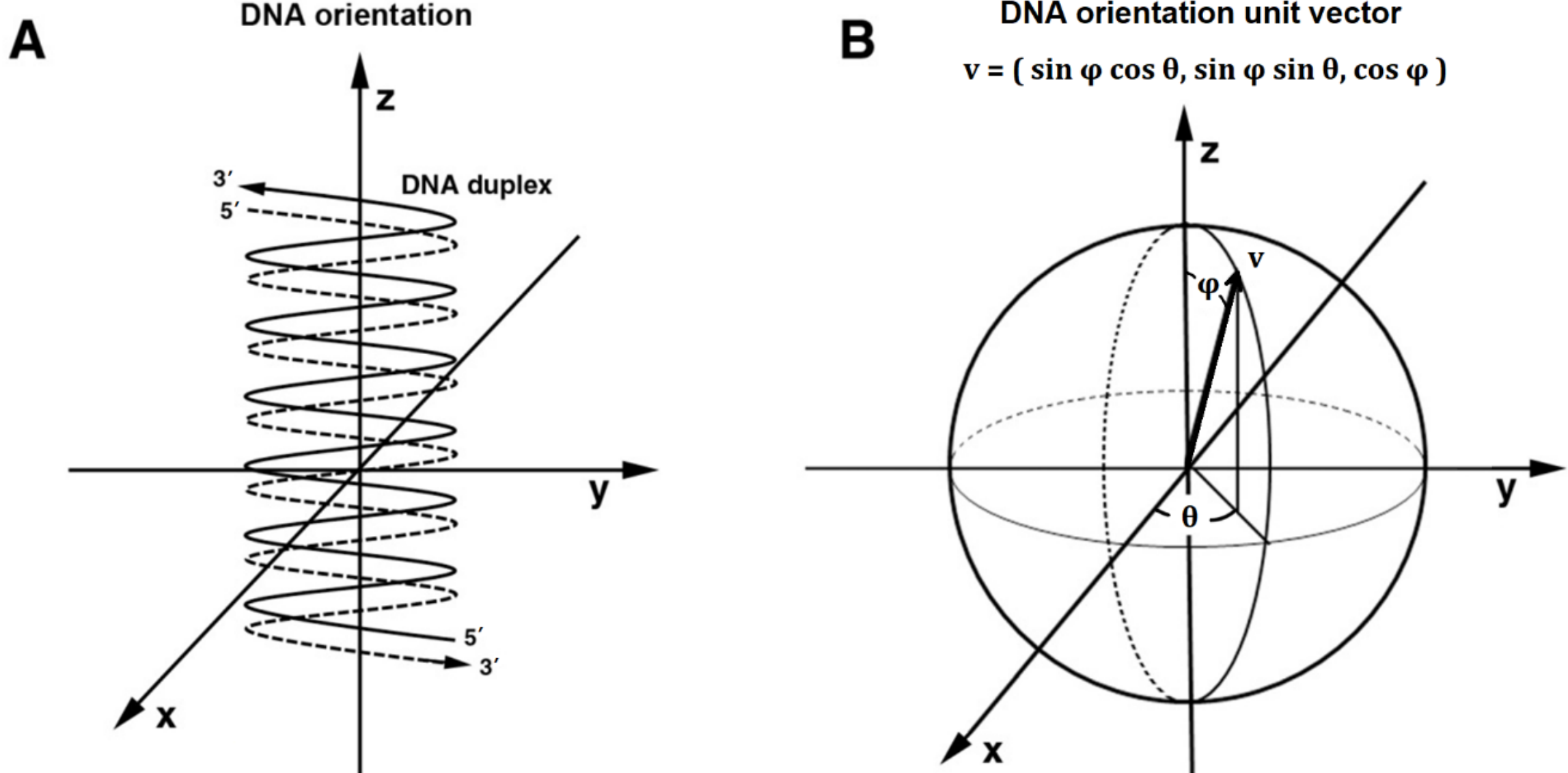

**Figure. 8.** A model of the physical efficiency of CEDS on dsDNAs. (A) A dsDNA oriented in the +z axis direction with sense sequence (solid line) pointing upwards has the full physical efficiency 1 with respect to its corresponding CEDS. A dsDNA oriented in the -z axis direction has the physical efficiency 0 with respect to the same CEDS. (B) When the physical efficiency of CEDS on a dsDNA oriented in the (θ,φ) direction is modeled as $\eta(\theta,\varphi) = \begin{cases} \cos\varphi, & 0 < \varphi < \pi/2 \\ 0, & \frac{\pi}{2} \leq \varphi < \pi \end{cases}$, the average physical efficiency of CEDS on randomly distributed dsDNAs in three dimensions is given by 25%.

**CEDS effect on electrical induction in ds3T3A and ds3C3G, which consist of only A-T and G-C base pairs, respectively.**

According to the HBMR hypothesis, the repeated CEDS can accumulate the electrical charge in the purine and pyrimidine bases of target oligo-dsDNAs. However, the accumulated electrical charge in the oligo-dsDNAs will be slowly discharged into the buffer solution. To accelerate the electrical discharge, an auxiliary electric current between the oligo-dsDNA molecules has been installed using an electric current detection device (see Figure 9 A), and then a little amount of electrical discharge from oligo-dsDNAs is able to be measured.

Two electrical probes, each approximately 1 mm in diameter, were placed 0.5 mm apart in a 1.5 mL tube. The auxiliary electric current was approximately 0.5 μA AC, with an absolute sinusoidal wave at 120 Hz, between the probes. The oligo-dsDNA sample, dissolved in a 0.1 M NaCl solution, was added to the tube. After CEDS using 13-15 Gauss magnetic field for 30 min, the auxiliary electric current was connected, and then the electric current (μA) discharged from the oligo-dsDNAs was measured between the electric probes using a precision electric meter, Fluke 289 (USA, Everett), for 90 min.

The ds3T3A (100 pmole/μL) in 0.1 M NaCl solution was treated with 3T3A*-CEDS using a 13-15 Gauss magnetic field for 30 min. The electric current was then measured under the stimulation of an auxiliary electric current by applying 0.5 μA absolute sine wave at 120 Hz during the resting period for 90 min. It was observed that the electric current increased significantly at 280 msec exposure time for A-T base pair during the 40-70 minute experimental period (Fig. 9 C). On the other hand, the ds3T3A in 0.1 M NaCl solution was not treated with CEDS for 30 min, there was no significant increase of electric current during the resting time for 90 min (Fig. 9 B). These data may indicate that CEDS effect on dsDNA could be evaluated by measuring the electric current discharged from dsDNA after CEDS treatment.

These data suggested that the increase in electric current discharged from oligo-dsDNAs after CEDS may be derived from the accumulated electrical charge in dsDNA by CEDS, and partially represents CEDS effect on oligo-dsDNAs. Therefore, the comparison between the electric current discharge patterns after 3T3A*-CEDS or 3C3G*-CEDS could be used to determine the effective exposure time of magnetic field for A-T and G-C pairs, respectively.

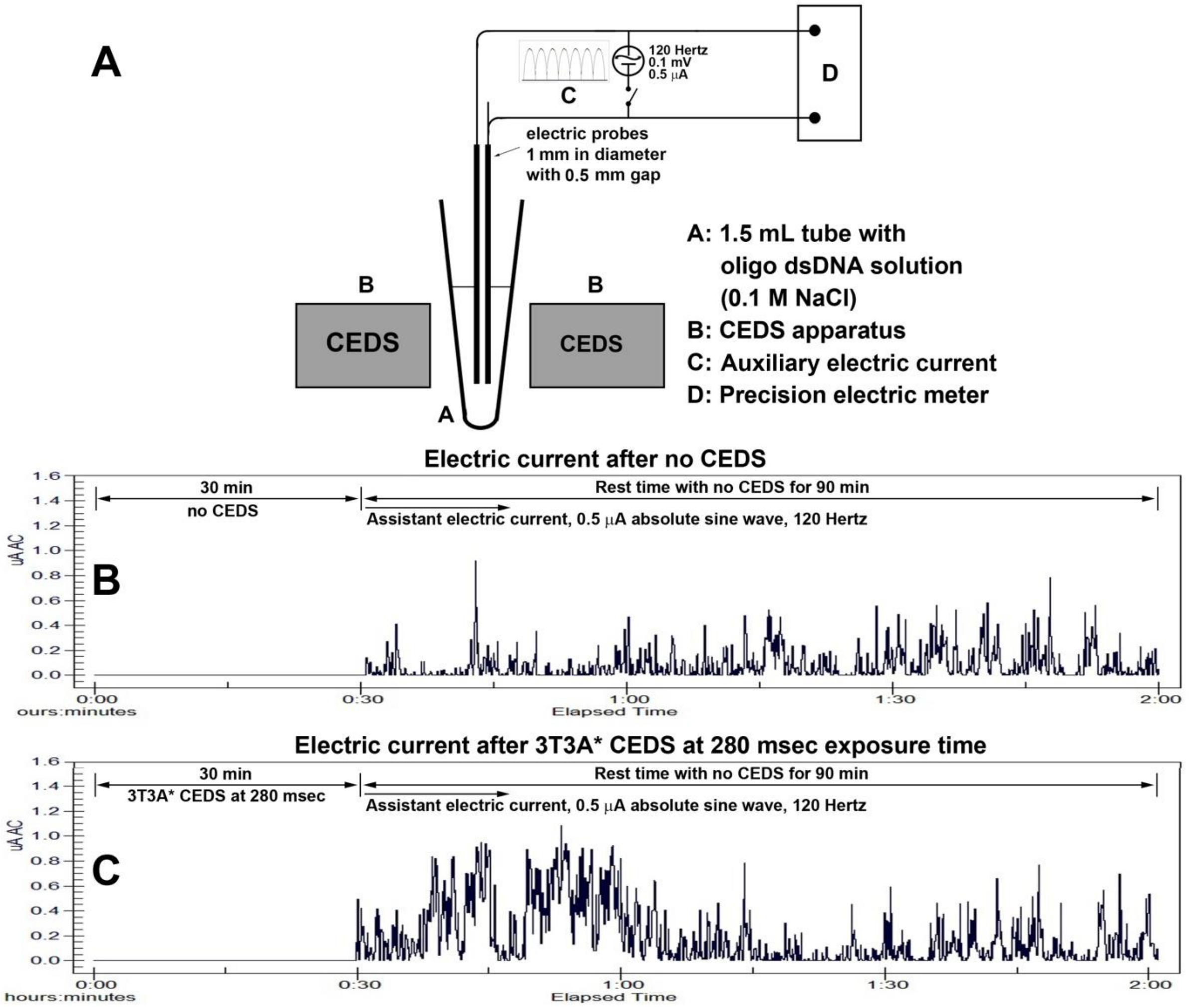

**Figure 9.** Detection of electric current discharged from oligo-dsDNAs after CEDS treatment. A: Schematic illustration of electric current detection apparatus installed with auxiliary electric current, 0.5 µA absolute sine wave 120 Hertz, between electrodes. The increase of electric current after 3T3A*-CEDS on ds3T3A in 0.1 M NaCl solution with 12-13 Gauss magnetic field at 280 msec exposure time per A-T base pairs for 30 min was found at 40-70 min of experiment during the resting time for 90 min (B), while no increase of electric current after no CEDS treatment (C).

In order to determine the optimal magnetic exposure time on the A-T and G-C base pairs of dsDNA, a series of experiments were performed using the same methods as above to compare the amount of electric current discharge from ds3T3A and ds3C3G after 3T3A*-CEDS and 3G3C*-CEDS, respectively, at different magnetic exposure times, 160, 200, 240, 280, 320, 360, 400, 440, 480, 520, 560, and 600 msec. As a result, ds3T3A, which is composed of only A-T base pairs, showed the dominant increase of electric current discharge at 240 -280 msec magnetic exposure time compared to ds3C3G, which is composed of only G-C base pairs. On the other hand, ds3C3G showed the dominant increase of the releasing electric current at 440-480 msec magnetic exposure time compared to ds3T3A (Fig. 10).

These data suggest that CEDS effect on the A-T base pairs of dsDNA is more enhanced at the magnetic exposure time of 240-280 msec compared to CEDS effect on the G-C base pairs of dsDNA, and CEDS effect on the G-C base pairs of dsDNA is more enhanced at the magnetic exposure time of 440-480 msec compared to CEDS effect on the A-T base pairs of dsDNA. Therefore, in the present CEDS experiments, every A-T base pairs and G-C base pairs of dsDNAs were exposed for 280 msec and 480 msec, respectively.

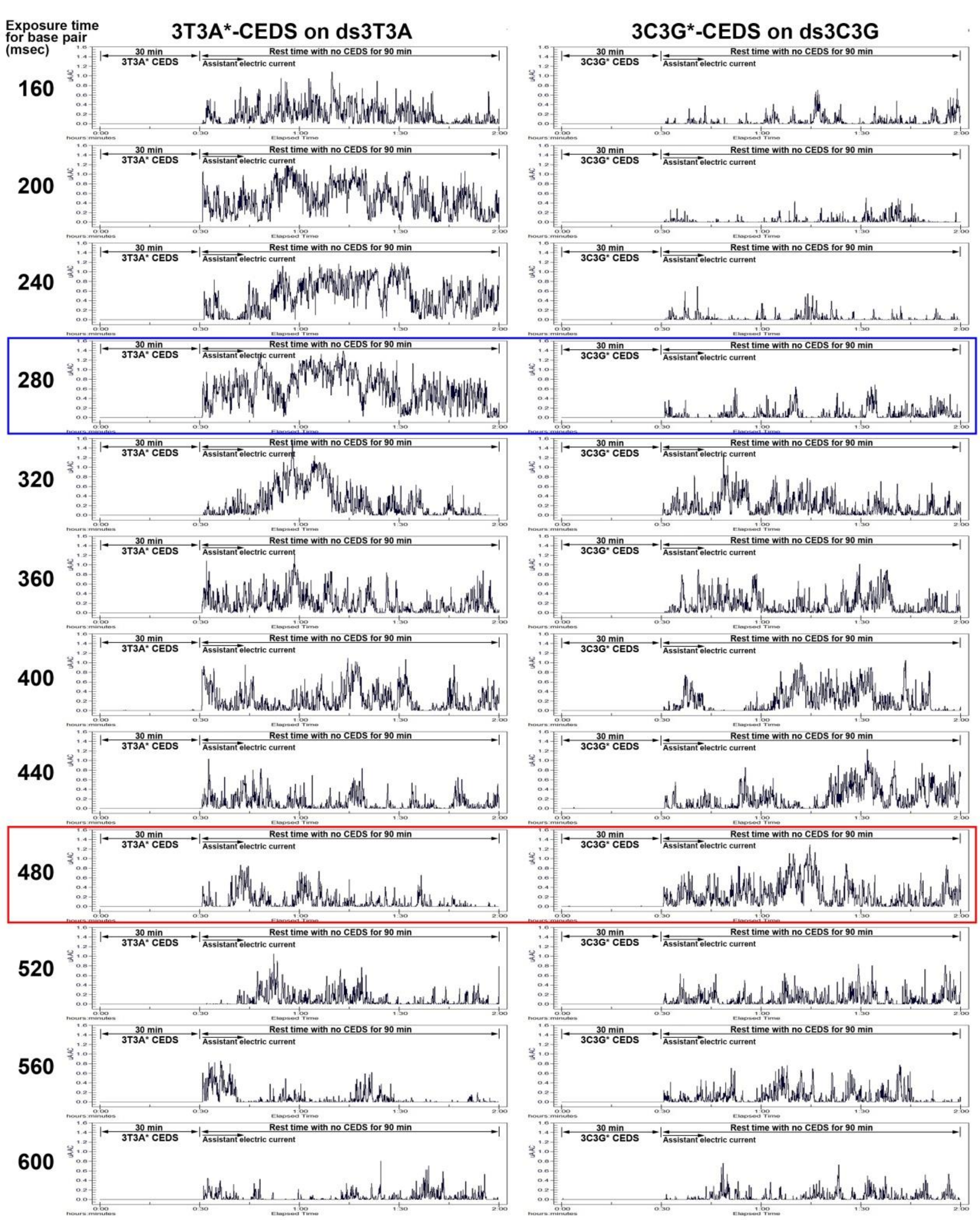

**Figure 10.** Comparison of electric induction potential in oligo-dsDNAs by CEDS at different magnetic exposure time, 160-600 msec, between A-T base pairs of ds3T3A and G-C base pairs of ds3C3G. Noted the most dominant increase of electric current in ds3T3A at 280 msec magnetic exposure time contrast in ds3C3G (blue square), and the most dominant increase of electric current in ds3C3G at 480 msec magnetic exposure time contrast in ds3T3A (red square).

**CEDS effect on the hybridization of oligo-dsDNAs**

***DNA hybridization assay in 0.005M NaCl solution through HPLC analysis***

The use of 0.1M or 0.2M NaCl solutions in the HPLC assay can cause spontaneous hybridization of ssDNA pairs, making it difficult to distinguish the real CEDS effect on DNA hybridization. To ensure accurate results in the DNA hybridization assay, we first determined the optimal NaCl concentration for achieving equilibrium between oligo-dsDNAs and oligo-ssDNAs. This was done by decreasing the NaCl concentration in the HPLC-based hybridization assay.

The two DNA molecules, ss2(3C3A) and ss2(3T3G), were dissolved at a concentration of 100 pmole/µL separately in NaCl solutions of varying concentrations, 0.2M, 0.05M, 0.01M, and 0.005M. After 10 min at room temperature, each mixture was analyzed using HPLC. The analysis was performed using a reverse-phase silica column with a mobile phase of NaCl at the different concentrations, 0.2M, 0.05M, 0.01M, and 0.005M at a flow rate of 0.3 mL/min and monitored at UV260 nm.

At room temperature, a pair of ss2(3C3A) and ss2(3T3G) hybridized spontaneously in 0.2M and 0.05M NaCl solution, producing dominant dsDNA peaks. However, when the pair of oligo-ssDNAs hybridized in 0.01M and 0.005M NaCl solution, they partially hybridized and produced two dominant peaks of dsDNAs (shorter retention time) and ssDNAs (longer retention time). Specifically, in 0.005M NaCl solution, the pair of oligo-ssDNAs showed an almost equilibrium state between dsDNAs and ssDNAs (Fig. 11 A).

Meanwhile, ss2(3C3A) and ss2(3T3G) were dissolved in 0.005 M NaCl solutions and heated to 40℃, 60℃, and 90℃, then gradually cooled to room temperature separately. After standing at room temperature for 10 min, they were analyzed by HPLC as described above. The results showed a relatively even equilibrium state of dsDNAs and ssDNAs at 60℃ compared to 40℃ and 90℃ (Fig. 11 B).

To investigate CEDS effect on DNA hybridization, we conducted a new experiment using a pair of ss2(3C3A) and ss2(3T3G) dissolved in a 0.005M NaCl solution and annealed at 60℃. Then, the ssDNA pair was treated with 2(3C3A)*-CEDS for 0, 30, 60, 90, 120, 150, 180, and 210 min, followed by HPLC analysis as mentioned above.

CEDS resulted in a shift in the balance between double-stranded DNAs (dsDNAs) and single-stranded DNAs (ssDNAs), with a prominent peak area of dsDNAs and a small peak area of ssDNAs. The shift in the balance between double-stranded DNAs (dsDNAs) and single-stranded DNAs (ssDNAs) was dependent on CEDS time, as shown in Figure 11 C. A plotted graph of the data clearly shows a gradual increase in the dsDNA peak for ds2(3C3A), and a gradual decrease in the ssDNA peak for ss2(3C3A) and ss(3T3G), also dependent on CEDS time (Fig. 11 D). The data indicate that 2(3C3A)*-CEDS induced hybridization of the oligo-ssDNA pair, ss2(3C3A) and ss2(3T3G), dissolved in 0.005M NaCl solution, resulted in a dominant increase of the ds2(3C3A) peak depending on CEDS time compared to the ss2(3C3A)/ss2(3T3G) peak.

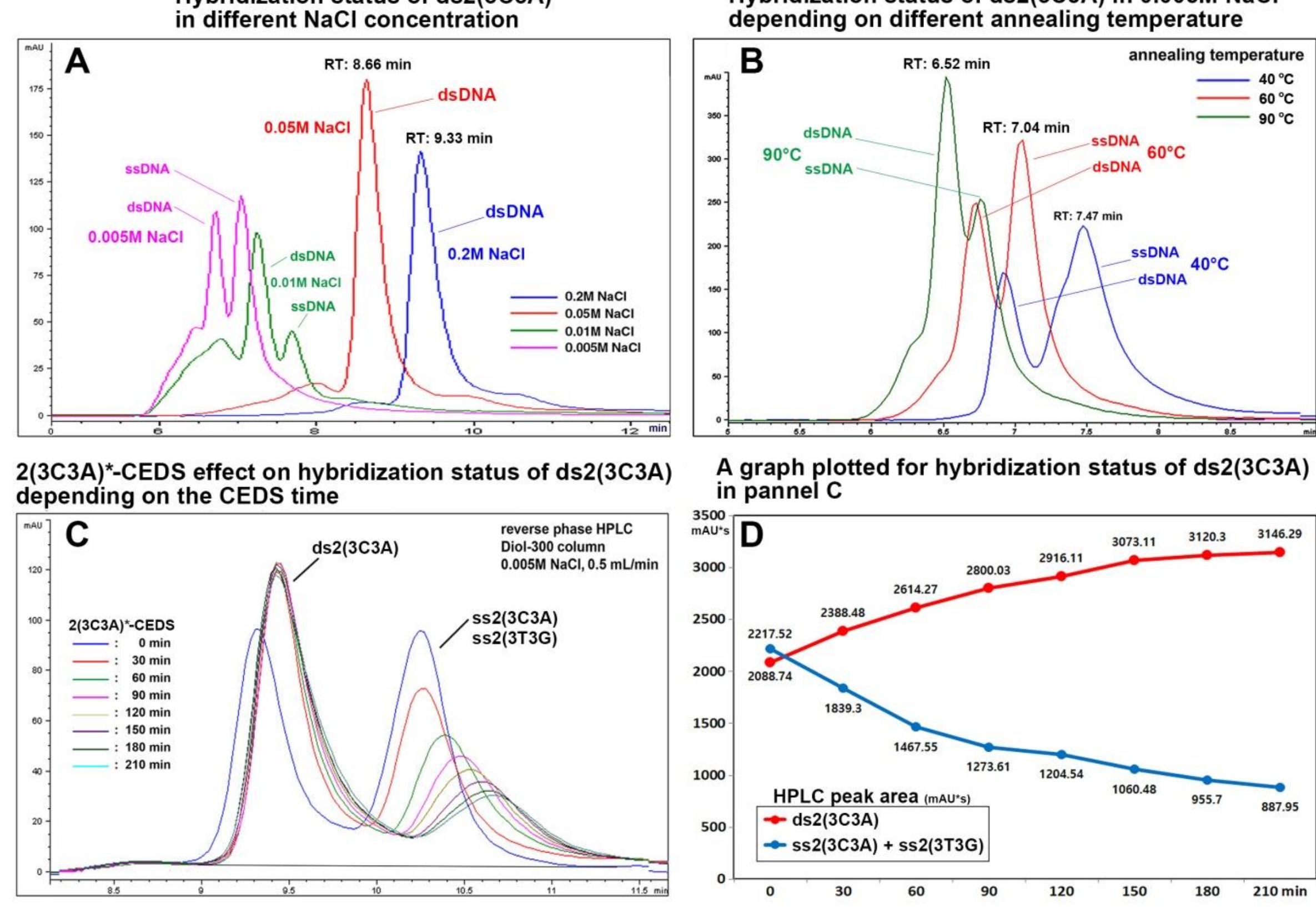

**Figure 11.** HPLC analysis of oligo-Pyu dsDNA hybridization status depending on NaCl concentration. A pair of oligo-ssDNAs, ss2(3C3A) and ss2(3T3G), was strongly hybridized in 0.05M and 0.2M NaCl solutions at room temperature, while loosely hybridized in 0.005M and 0.01M NaCl solutions. In particular, the oligo-ssDNA pairs showed an almost equilibrium state of oligo-dsDNAs and oligo-ssDNAs in 0.005M NaCl solution at room temperature (A). This state of equilibrium was most prominent when annealed at 60℃ compared to 40℃ and 90℃ (B). To investigate CEDS effect on DNA hybridization, a new experiment was conducted utilizing a pair of ss2(3C3A) and ss2(3T3G) that were dissolved in a 0.005M NaCl solution and annealed at 60℃ through HPLC. The ssDNA pairing was then hybridized by 2(3C3A)*-CEDS depending on CEDS time, 30, 60, 90, 120, 150, 180, and 210 min (C). A graph was created using panel C data, showing a gradual increase in ds2(3C3A) peak and a gradual decrease in ss2(3C3A) and ss(3T3G) peaks (D). RT: Retention time.

***2) CEDS-induced hybridization of oligo-ssDNA pairs is reversible within 18 hours after CEDS***

To investigate the effect of CEDS on DNA hybridization in the partially hybridized state of oligo-dsDNA, various Pyu oligo-ssDNA pairs, such as ss2(6C6A) and ss2(6T6G), ss3(4C4A) and ss3(4T4G), ss4(3C3A) and ss4(3T3G), ss6(2C2A) and ss6(2T2G), ss12(CA) and ss12(TG), and ss12C12A and ss2T12G were dissolved in 0. 005M NaCl solution. Each Pyu oligo-ssDNA pair partially hybridized in 0.005M solution at room temperature spontaneously, and then treated with CEDS using a corresponding sequence at 20-25 Gauss for 30 min.

The DNA hybridization status of the sample was analyzed using HPLC with a non-adherent silica column and 0.005M NaCl solution mobile buffer at a flow rate of 0.3 mL/min and UV260 nm. Each experimental result was compared to CEDS-untreated control. Additionally, to determine the recovery of CEDS effect, CEDS-treated oligo-DNA samples were kept at room temperature for 18 hours, and then each sample was analyzed by HPLC using the same procedures as described above.

As a result, when the oligo-ssDNA pairs for Pyu oligo-dsDNAs, ds3(4C4A), ds4(3C3A), ds6(2C2A), and ds12(CA), were dissolved in 0.005M NaCl solution, they were hybridized more by CEDS using a corresponding sequence than the untreated controls (Fig. 12 C, E, G, and I), and these DNA hybridizations were recovered to the normal state of the untreated controls in 18 hours at room temperature (Fig. 12 D, F, H, and J). On the other hand, the oligo-ssDNA pairs for Pyu oligo-dsDNAs, ds2(6C6A) and ds12C12A, appeared to hybridize spontaneously but incompletely in a 0.005M NaCl solution. The samples were slightly more hybridized by CEDS than the untreated controls (Fig. 12 A, B, K, and L). However, it was postulated that ds2(6C6A) and ds12C12A underwent atypical hybridization with multiple G-quadruplex complexes, thereby exhibiting minimal change by CEDS.

The data show that CEDS-induced hybridization of ds3(4C4A), ds4(3C3A), ds6(2C2A), and ds12(CA) could be distinguished in a low salt solution of 0.005M NaCl, which ultimately returned to its normal state within 18 hours.

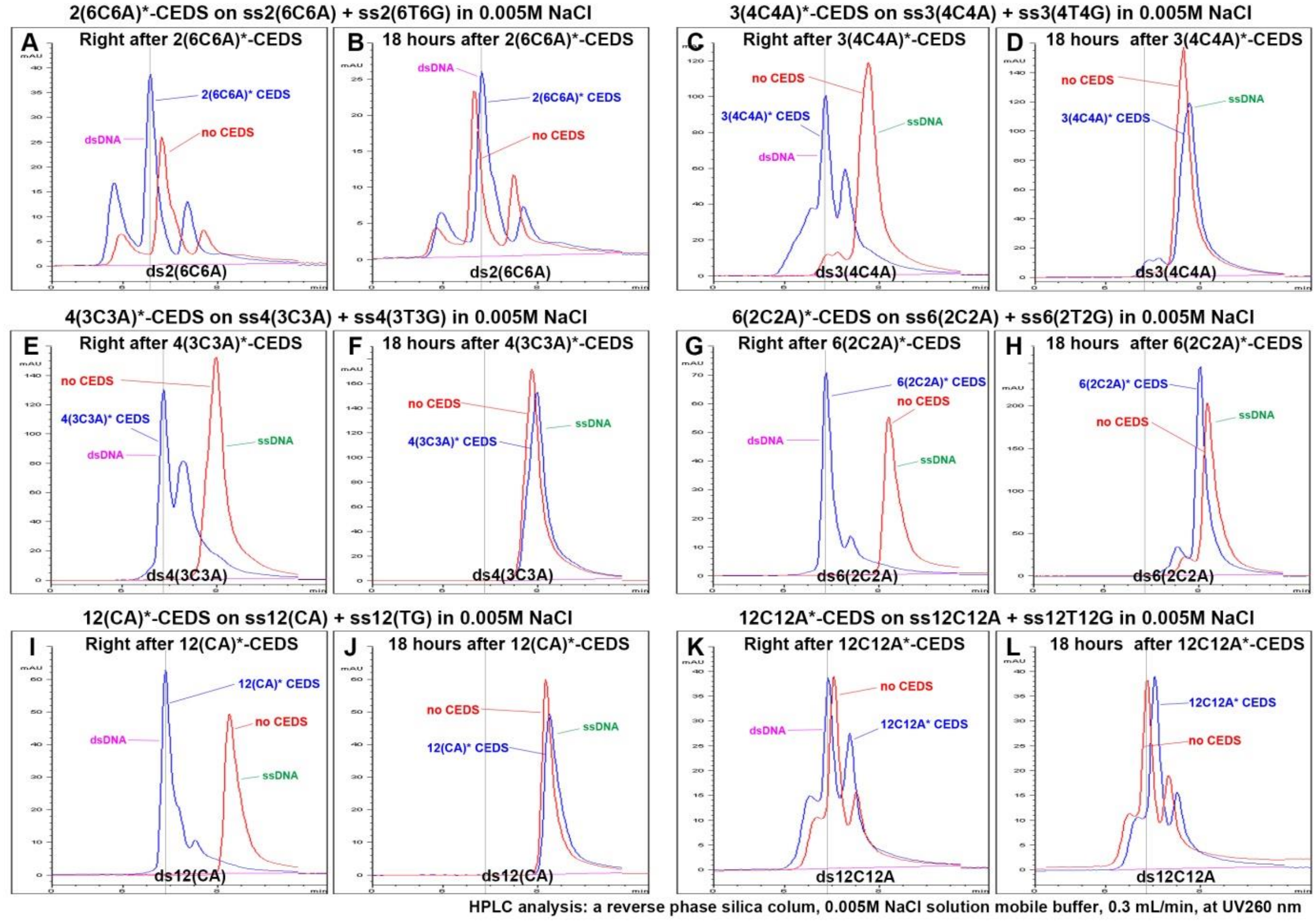

**Figure 12.** HPLC-based DNA hybridization assay in 0.005M NaCl solution. Pyu ssDNA pairs composed of short Pyu dsDNA segments, ss3(4C4A) and ss3(4T4G), ss4(3C3A) and ss4(3T3G), ss6(2C2A) and ss6(2T2G), and ss12C12A and ss12T12G, were consistently hybridized by CEDS using a corresponding sequence (C, E, G, and I), whereas the Pyu ssDNA pairs composed of long Pyu dsDNA segment(s), ss12C12A and ss12T12G, ss2(6C6A) and ss2(6T6G), hybridized spontaneously but incompletely, thus CEDS-induced hybridization appeared to be minimal (A and K). When the DNA sample treated with CEDS was kept free at room temperature for 18 hours, the oligo-dsDNAs hybridized by CEDS were almost converted to the normal state (B, D, F, H, J, and L).

***3) DNA hybridization of oligo-ssDNA pairs by CEDS is in a sequence-specific manner***

In order to know the specificity of CEDS targeting, the above low salt (0.005M NaCl) hybridization assay was performed using the ssDNA pairs of ss3(4C4A) and ss3(4T4G), ss4(3C3A) and ss4(3T3G), ss6(2C2A) and ss6(2T2G), ss12(CA) and ss12(TG) dissolved in 0.005M NaCl solution, excluding the multiple G-quadruplex complex-suspected dsDNAs, ds2(6C6A) and ds12C12A. The DNA hybridizations induced by CEDS using a target sequence for ds3(4C4A), ds4(3C3A), ds6(2C2A), ds12(CA), or a random sequence ds3(ACGT) were assessed by HPLC analysis.

The results showed that the target sequence-CEDSs consistently induced dominant peaks of DNA hybridization in the sequence-specific patterns of ds3(4C4A), ds4(3C3A), ds6(2C2A), and ds12(CA) compared to the untreated controls (Fig. 13 A, C, E, and G), whereas the random sequence, 3(ACGT)*-CEDS did not induce a *de novo* peak of those DNA hybridization (Fig. 13 B, D, F, and H). These data suggest that the ssDNA pairs ss3(4C4A) and ss3(4T4G), ss4(3C3A) and ss4(3T3G), ss6(2C2A) and ss6(2T2G), ss12(CA) and ss12(TG) were specifically hybridized by the target sequence*-CEDSs, whereas they were rarely hybridized by the random sequence, 3(ACGT)*-CEDS.

Taken together, it was evident that the target sequence*-CEDSs have a hybridization potential of target oligo-ssDNA pairs dissolved in 0.005M NaCl solution depending on CEDS time up to 210 min, and in the sequence-specific manner, when compared to the random sequence, 3(ACGT)*-CEDS. Additionally, CEDS-induced hybridization of oligo-ssDNA pairs was reversible to the normal state within 18 hours after CEDS treatment.

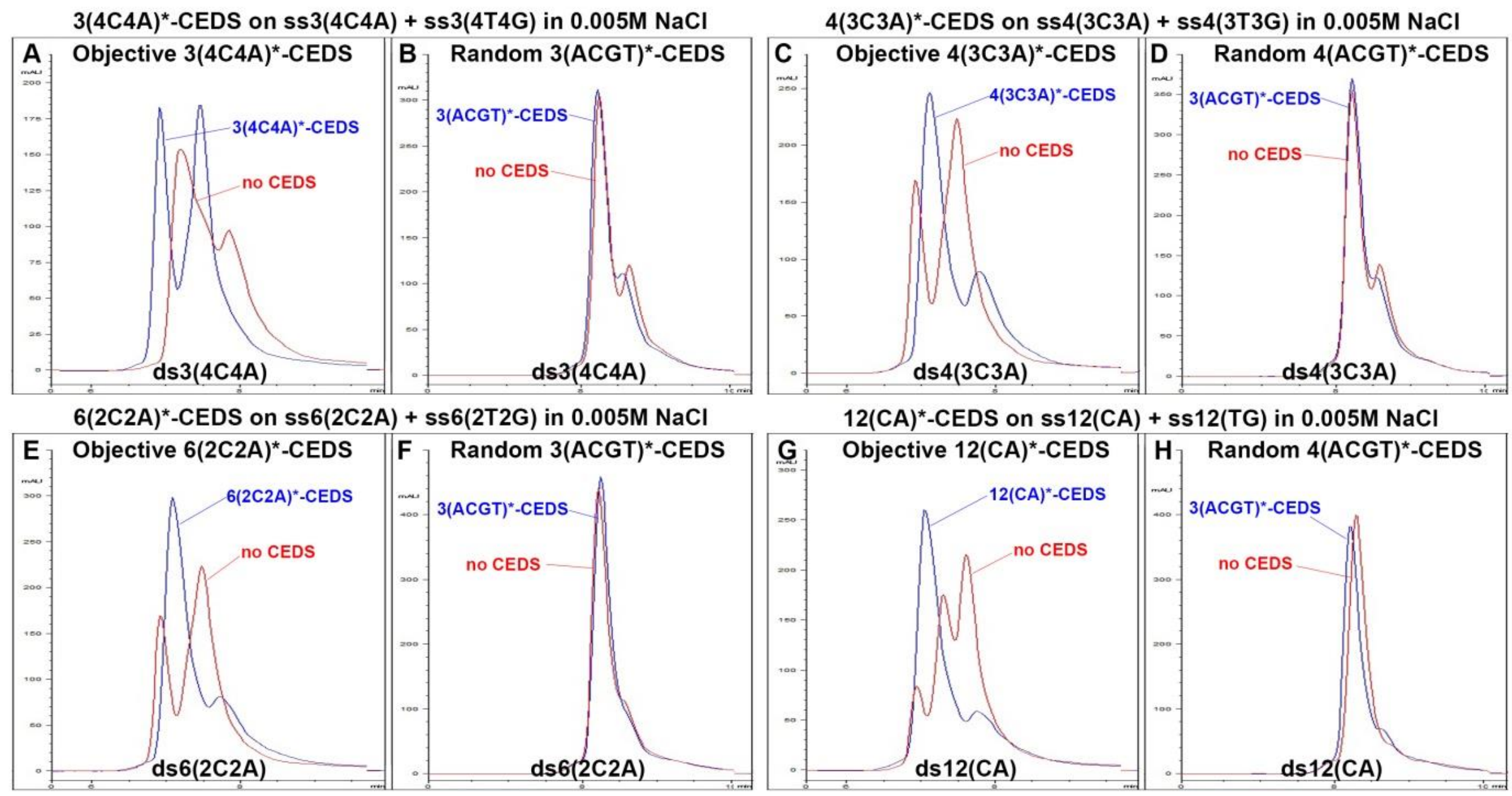

**Figure 13.** HPLC-based DNA hybridization assay in 0.005M NaCl solution. Pyu oligo-ssDNA pairs for Pyu oligo-dsDNAs, ss3(4C4A) and ss3(4T4G), ss4(3C3A) and ss4(3T3G), ss6(2C2A) and ss6(2T2G), and ss12(CA) and ss12(TG), were specifically hybridized by 3(4C4A)*-CEDS, 4(3C3A)*-CEDS, 6(2C2A)*-CEDS, and 12(CA)*-CEDS, respectively (A, C, E, and G), compared to the control treated with random sequence-CEDS, 3(ACGT)*-CEDS (B, D, F, and H).

**CEDS effect on the DNA conformation of oligo-dsDNAs**

The dsDNAs must change their three-dimensional conformation to transcribe RNAs and DNAs by interacting with various proteins. Identifying the exact conformational changes of long and coiled dsDNA can be very challenging. Therefore, this study used different Pyu and Puy oligo-dsDNAs to investigate the effect of CEDS on the conformational changes of dsDNA in 0.1M NaCl solution. It is expected that CEDS can induce the morphological changes of oligo-dsDNAs, i.e. elongation or contraction, swelling or constriction, etc., but not the structural and chemical changes. These morphological changes may partly represent the conformational changes of dsDNAs on a three-dimensional scale and are detectable by UV spectroscopy of HPLC at 260 nm.

Six nucleotide-long oligo-ssDNAs were purchased from Cosmogenetech Co. Ltd. (Seoul, Korea). Each nucleotide, including ss3T3A, ss3A3T, ss3C3G, and ss3G3C, was separately dissolved in 0.1 M NaCl solution at 100 pmole/μL to produce self-dimeric hybridization of ds3T3A, ds3A3T, ds3C3G, and ds3G3C after heating to 95℃ and slow cooling at room temperature. Furthermore, to create ds3C3A and ds3A3C, two pairs of ss3C3A and ss3T3G, ss3A3C and ss3G3T were hybridized by heating to 95℃ and slow cooling at room temperature separately. The resulting Pyu and Puy oligo-dsDNAs are expected to be free of tertiary conformational changes. The dsDNA samples were treated with CEDS using a corresponding sequence for 20 min separately. Subsequently, the samples were subjected to analysis via HPLC with a reverse-phase silica column and a running buffer of 0.1 M NaCl solution at a flow rate of 0.3 mL/min.

For oligo-dsDNAs comprising six sequences, the Pyu dsDNA, ds3T3A demonstrated a 4.8% increase in conformational change by 3T3A*-CEDS, while the Puy dsDNA, ds3A3T exhibited a 0.8% decrease by 3A3T*-CEDS. In contrast, 12A*-CEDS resulted in a 6.9% increase in the conformational size of ds3T3A and a 2.2% increase in the conformational size of ds3A3T. Moreover, the Pyu dsDNA, ds3C3G demonstrated a 5.2% increase in conformational change by 3C3G*-CEDS, while the Puy dsDNA, ds3G3C exhibited a 10.4% increase by 3G3C*-CEDS. In contrast, 12A*-CEDS resulted in a 12.1% increase in the conformational size of ds3C3G and a 13.1% increase in the conformational size of ds3G3C. Furthermore, the Pyu dsDNA, ds3C3A exhibited a 0.7% reduction in conformational change by 3C3A*-CEDS, while the Puy dsDNA, ds3A3C demonstrated a 2.2% reduction by 3A3C*-CEDS. In contrast, 12A*-CEDS resulted in a 0.9% reduction in the conformational size of ds3C3A and a 7.5% reduction in the conformational size of ds3A3C.

The data indicate that oligo-dsDNAs exhibit unique conformational changes by CEDS using a corresponding sequence in a manner that differs between Pyu and Puy dsDNAs and between A-T and G-C pairs. As a result, 3T3A*-CEDS increased the conformational size of ds3T3A by 4.8%, while 3A3T*-CEDS decreased the conformational size of ds3A3T by 0.8%. 3C3G*-CEDS increased the conformational size of ds3C3G by 5.2%, while 3G3C*-CEDS increased the conformational size of ds3G3C by 10.4%. And 3C3A*-CEDS decreased the conformational size of ds3C3A by 0.7%, while 3A3C*-CEDS decreased the conformational size of ds3A3C by 2.2%.

This study compared the conformational changes of oligo-dsDNA induced by CEDS using a target sequence with those using a nonspecific sequence (12A). 12A*-CEDS may produce a unipolar and monotonous cyclic magnetic field that can have either positive or negative effects on the entire sequence of target dsDNAs. This can ultimately result in an abnormal DNA conformation that may be unsuitable for the normal functions of dsDNA.

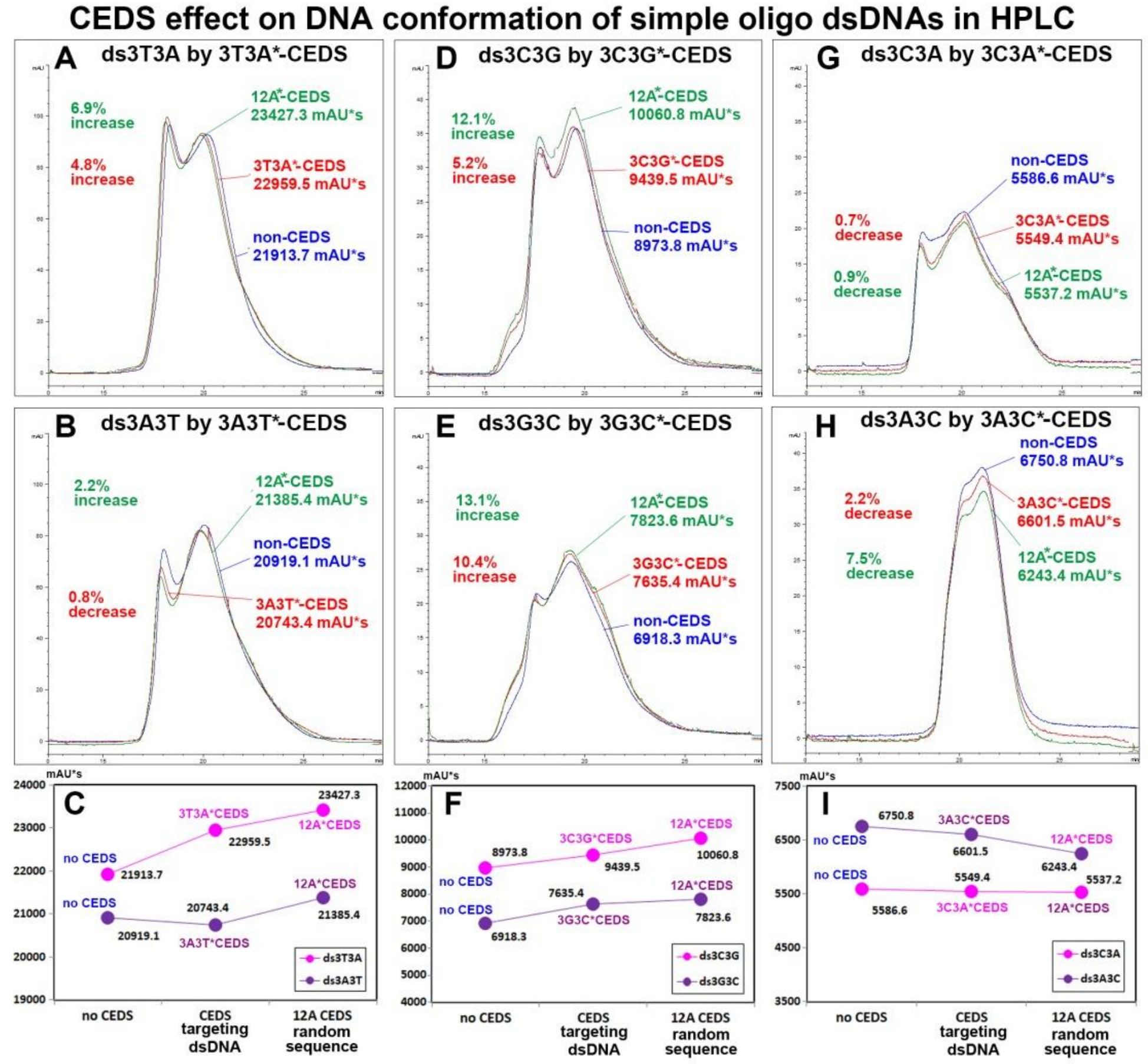

**Figure 14.** CEDS-induced conformational changes of simple six sequence dsDNAs were assessed by HPLC analysis. A: Pyu dsDNA, ds3T3A by 3T3A*-CEDS. B: Puy dsDNA, ds3A3T by the 3A3T*CEDS. D: Pyu dsDNA, ds3C3G by 3C3G*-CEDS. E: Puy dsDNA, ds3G3C by 3G3C*-CEDS. G: Pyu dsDNA, ds3C3A by 3C3A*-CEDS. H: Puy dsDNA, ds3A3C by 3A3C*-CEDS. C: A graph for ds3T3A versus ds3A3T. F: A graph for ds3C3G versus ds3C3G. I: A graph for ds3C3A versus ds3A3C. Blue line: the untreated control. Red line: CEDS using a target sequence. Green line: nonspecific sequence, 12A-CEDS.

CEDS effect on DNA conformation was also investigated for different DNA motif sequences, restriction endonuclease (RE) binding sites, canonical E-box and telomere by HPLC. The RE binding sites, XhoI, SacI, AflII, EcoRI, XbaI, BglII, BstBI, and HindIII, are palindromic six base pairs dsDNAs composed of and a Pyu or Puy dsDNA segment(s), canonical E-box is a palindromic six base pairs dsDNA composed of three Pyu dsDNA segments, and the telomere is a six base pairs dsDNA composed of a Pyu dsDNA segment. The oligo-dsDNAs of DNA motif sequences were generated and examined for their conformational changes after CEDS using each corresponding sequence through the same procedures previously described.

For the oligo-dsDNAs consisting of complex six sequences for different endonuclease restriction sites, dsCTCGAG, a palindromic Pyu sequence for XhoI restriction site, showed 14.8% decrease in conformational change by CTCGAG*-CEDS and 14. 1% decrease by 12A*-CEDS compared to the untreated control, whereas dsGAGCTC, a palindromic Pyu sequence for XhoI restriction site, showed 12.9% decrease in conformational change by GAGCTC*-CEDS and 15.6% increase by 12A*-CEDS (Fig. 15 A and B).

The dsCTTAAG is a palindromic Pyu sequence that contains a XhoI restriction site. When treated with CTTAAG*-CEDS, it showed a 13.2% increase in conformational change and a 4.2% decrease with the 12A*-CEDS. On the other hand, the dsGAATTC, which is also a palindromic Puy sequence with a XhoI restriction site, showed a 3% decrease with GAATTC*-CEDS and a 1.8% increase with 12A*-CEDS (Fig. 15 C and D).

The dsTCTAGA, a palindromic Pyu sequence for XhoI restriction site, showed 5.2% increase in conformational change by TCTAGA*-CEDS and 1.4% increase by 12A*-CEDS, whereas dsAGATCT, a palindromic Puy sequence for XhoI restriction site, showed 3% decrease by AGATCT*-CEDS and 21.2% decrease by 12A*-CEDS (Fig. 15 E and F).

The dsTTCGAA, a palindromic Pyu sequence for XhoI restriction site, showed 3.1% increase in conformational change by TTCGAA*-CEDS and 1.9% increase by 12A*-CEDS, whereas dsAAGCTT, a palindromic Puy sequence for XhoI restriction site, showed 6% decrease by AAGCTT*-CEDS and 9.4% decrease by 12A*-CEDS (Fig. 15 G and H).

And the dsTTAGGG, a telomere repeat sequence consisting of one Pyu dsDNA, showed 4% increase in conformational change by TTAGGG*-CEDS and 2.7% increase by 12A*-CEDS, whereas dsCACGTG, a canonical E-box sequence consisting of three Pyu dsDNA, showed 3.3% increase in conformational change by the CACGTG*-CEDS and 2.4% decrease by the 12A*-CEDS (Fig. 15 I and J).

The conformational changes of oligo-dsDNA induced by CEDS using a target sequence appear irregular in HPLC analysis. However, these changes are still unique compared to those induced by CEDS using a nonspecific sequence (12A). The total electric potential of palindromic oligo-dsDNAs encoding different RE binding site appears to be zero due to their balanced polarities between Pyu or Puy ssDNA segment(s), which can facilitate the formation of unique DNA conformations. It is therefore postulated that CEDS using a target sequence can induce unique conformations of the target oligo-dsDNAs encoding different RE binding sites. Conversely, CEDS using a nonspecific sequence (12A) may abnormally increase or decrease the conformation of oligo-dsDNAs in comparison to CEDS using a target sequence.

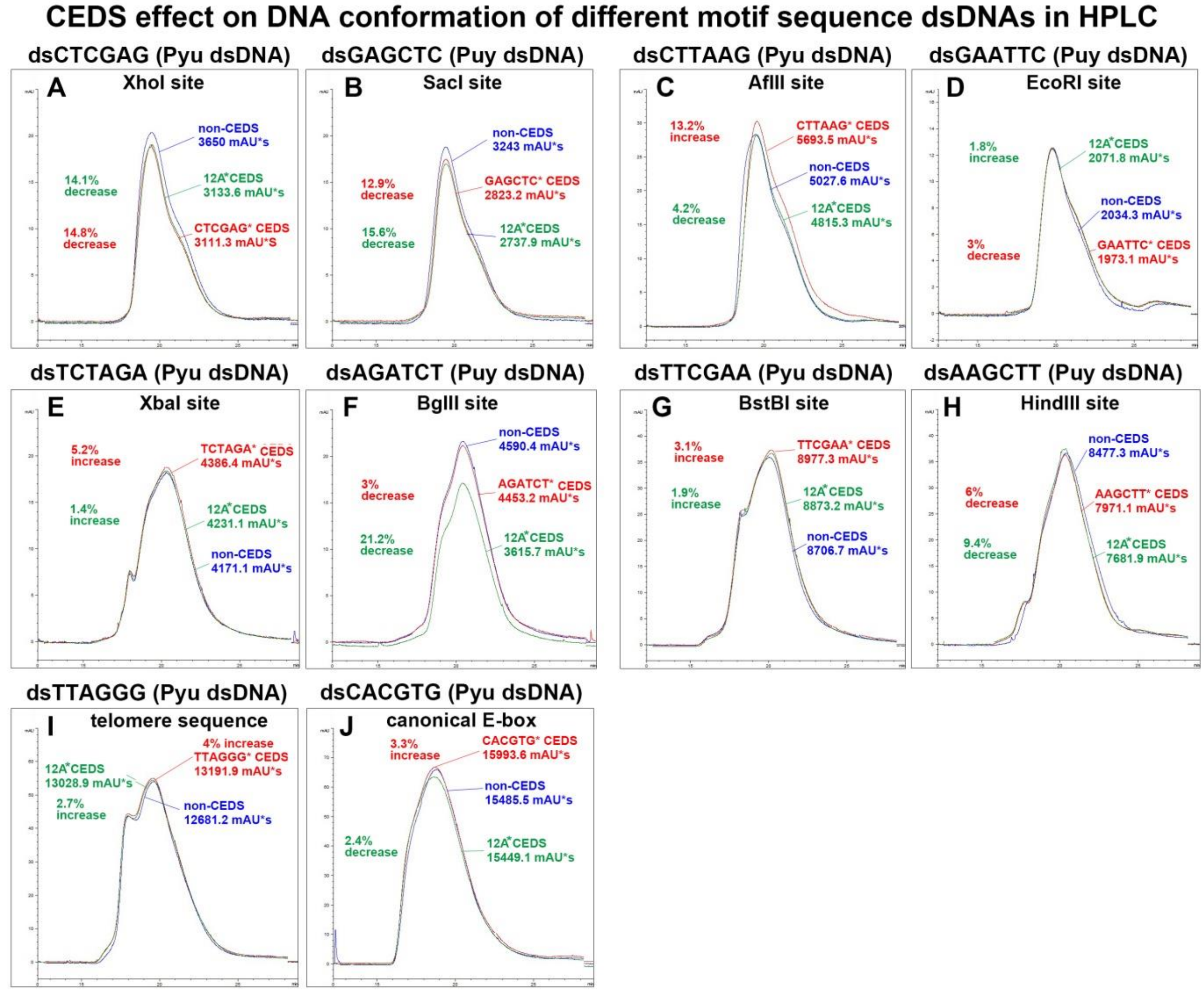

**Figure 15.** CEDS-induced conformational changes of complex six sequence dsDNAs encoding various RE binding sites were assessed by HPLC analysis. A and B: dsCTCGAG (XhoI) versus dsGAGCTC (SacI). C and D: dsCTTAAG (AflII) versus dsGAATTC (EcoRI). E and F: dsTCTAGA (XbaI) versus dsAGATCT (BglII). G and H: dsTTCGAA (BstBI) versus dsAAGCTT (HindIII). Blue line: CEDS untreated as a control. Red line: CEDS using a target sequence. Green line: CEDS using a nonspecific sequence (12A).

To determine the sequential conformational changes of Pyu oligo-dsDNA induced by CEDS, a 24-sequence-long Pyu dsDNA, ds6(2T2A), was separately treated with the target sequence (6(2T2A))*-, random sequence (6(ACGT))*-, and nonspecific sequence (24A)*-CEDS at room temperature for 15, 30, 45, 60, 75, 90, 105, and 120 min, and then analyzed by HPLC as previously described. 6(2T2A)*-CEDS demonstrated a consistent increase in the peak area of ds6(2T2A) with an increase in CEDS time, up to 120 min. Conversely, 6(ACGT)*-CEDS exhibited only a weak increase in the peak area of ds6(2T2A), while the 24A*-CEDS exhibited a slight decrease in the peak area of ds6(2T2A) during CEDS time (Fig. 16A).

The 24 sequence long Pyu dsDNA (6(2C2A)) was also treated separately with 6(2C2A)*-CEDS and 6(ACGT)*-CEDS for 90 min at room temperature. The samples treated with CEDS for 90 min were allowed to rest for 30, 60, 120, and 240 min before being analyzed using HPLC, as previously described. During CEDS period of 90 min, 6(2C2A)*-CEDS consistently increased the peak area of ds6(2C2A). The peak area of ds6(2C2A) exhibited a near-plateau during the resting period, with minimal change over the course of 120 min. Thereafter, the peak area of ds6(2C2A) exhibited a gradual decrease, reaching approximately half of its maximum increase value (51.8%) at 240 min. In contrast, 6(ACGT)*-CEDS produced irregular increases or decreases in the peak area of ds6(2C2A) of ds6(2C2A) during both CEDS and rest periods. The untreated control demonstrated minimal change in the peak area of ds6(2C2A) (Fig. 16 B).

The data indicates that both Pyu dsDNAs, ds6(2T2A) and ds6(2C2A), were consistently affected by 6(2T2A)*-CEDS and 6(2C2A)*-CEDS, respectively, in a sequence-specific manner, resulting in an increase of their HPLC peak areas during CEDS period. In particular, the increase in the peak area of ds6(2C2A) by 6(2C2A)*-CEDS was maintained until 120 min of the rest period, after which it gradually decreased almost to the half at 240 min of the rest period.

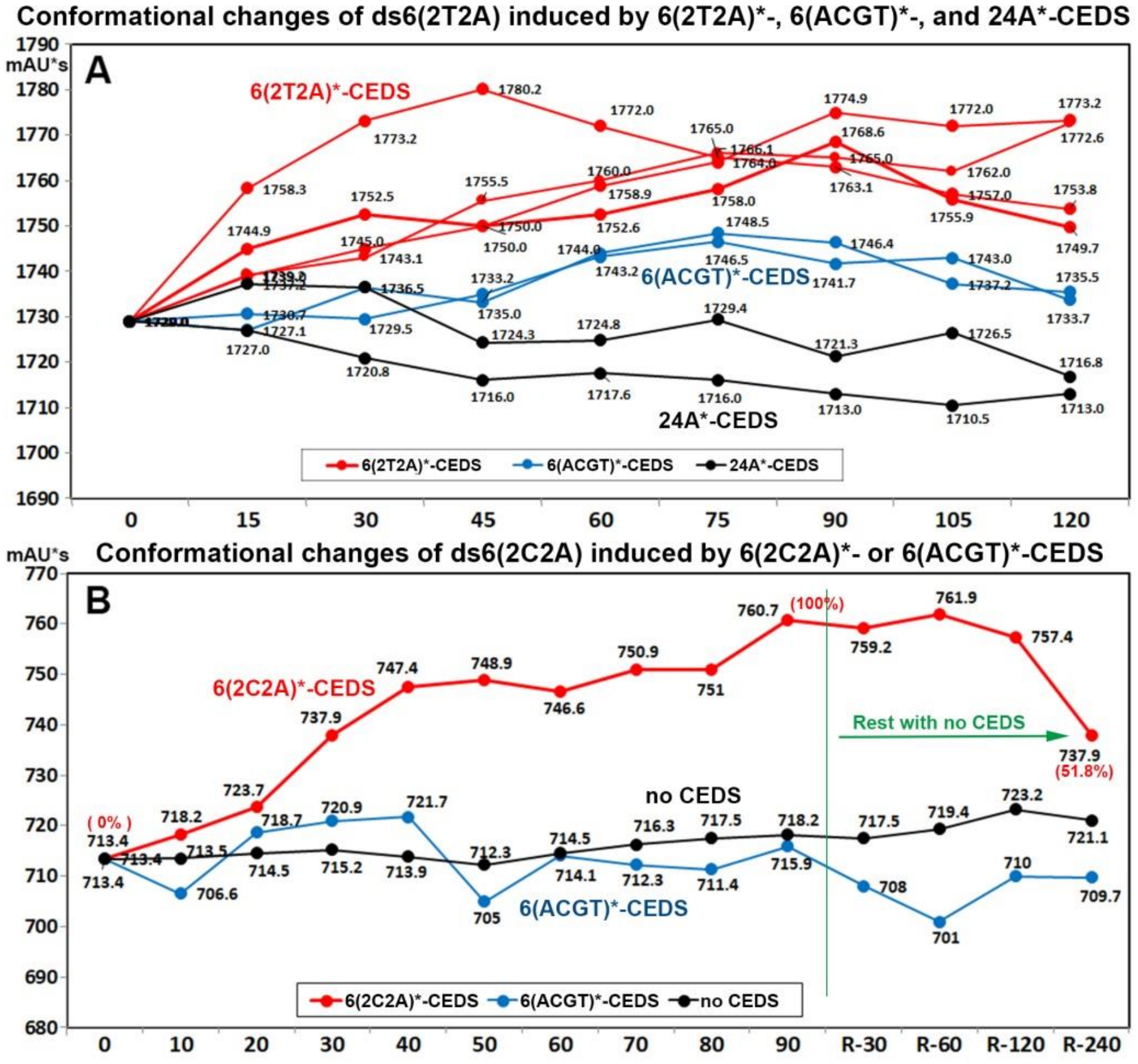

**Figure 16.** CEDS-induced conformational changes of 24 sequence-long dsDNAs through HPLC analysis. A: When the ds6(2T2A) was treated with 6(2T2A))*-CEDS, 6(ACGT)*-CEDS, and 24A*-CEDS, the 6(2T2A)*-CEDS markedly increased the peak area of ds6(2T2A) in HPLC depending on CEDS time until 120 min. While random sequence 6(ACGT)*-CEDS slightly increased the peak area of ds6(2T2A), and nonspecific sequence 24A*-CEDS slightly decreased the peak area of ds6(2T2A). B: When the ds6(2C2A) was treated with 6(2C2A)*-CEDS, 6(ACGT)*-CEDS, and 24A*-CEDS, the 6(2C2A)*-CEDS consistently increased the peak area of 6(2C2A), depending on CEDS time until 90 min. Thereafter, the peak area of ds6(2C2A)was maintained until 90 min of the rest period, but gradually decreased almost to the half at 240 min of the rest period. While 6(ACGT)* and 24A*-CEDS produced minimal changes of the peak area of ds6(2C2A).

**FT-IR changes of oligo-dsDNA by CEDS**

The oligo-dsDNAs in 0.1M NaCl solution may exist in the physiological DNA state, B- or Z-type DNA, supported by hydrogen bonds of water in ionic perturbation with Na+. The N-H stretching of the hydrogen bond donor site of oligo-dsDNA was detected by FT-IR spectroscopy at about 3700-2800 $cm^{-1}$ band centered at 3350-3300 $cm^{-1}$.

The simple oligo-dsDNAs consisting of only A-T or G-C base pairs, e.g. ds3T3A, ds3A3T, ds3C3G and ds3G3C, and the complex oligo-dsDNAs consisting of both A-T and G-C base pairs, such as ds3C3A, ds3A3C, dsCACGTG (canonical E-box sequence), dsTTAGGG (telomere repeat sequence), dsCTCGAG (XhoI binding site), dsGAGCTC (SacI binding site), dsCTTAAG (AflII binding site), dsGAATTC (BamHI binding site), and dsTCTAGA (XbaI binding site) were prepared at 100 pmole/μL using the annealing method described above. Each oligo-dsDNA was then treated with CEDS using a corresponding sequence at room temperature for 20 min. The samples were immediately analyzed using FT-IR (UATR Two, Perkin Elmer, USA). For the positive and negative controls, each oligo-dsDNA was treated with CEDS using a random sequence, 6(ACGT), and a nonspecific sequence (12A), following the same method described above.

In FT-IR analysis, to exclude the O-H stretching of water, FT-IR analysis was performed for the oligo-dsDNA samples together with DDW, 0.1M NaCl, and 1M NaCl solution as controls separately. Generally, the band at about 3700-2800 $cm^{-1}$ was higher in the 0.1M NaCl solution compared to DDW. The band increased significantly in the 1M NaCl solution and shifted to a higher frequency of wave. Generally, oligo-dsDNAs dissolved in 0.1M NaCl solution showed a slightly higher band than DDW, but a lower band than the 0.1M NaCl solution. However, oligo-dsDNA solution treated with CEDS using a corresponding sequence showed a significant increase in the infrared absorption band compared to untreated oligo-dsDNA solutions. The increase was still lower than that of the 0.1M NaCl solution. In contrast, oligo-dsDNA solutions treated with nonspecific sequence, 12A*-CEDS exhibited only a slight increase compared to the untreated oligo-dsDNA solution.

When comparing Pyu and Puy oligo-dsDNA solutions with ds3T3A and ds3A3T, ds3C3A and ds3A3C, dsCTTGAG and dsGAGCTC, dsTCTAGA and dsAGATCT, the infrared absorption bands of both Pyu and Puy dsDNAs were significantly increased by CEDS using a corresponding sequence, but weakly increased by CEDS using a nonspecific sequence (12A) compared to the untreated oligo-dsDNA solutions (Fig. 17 A-C, E, F, I-K, M, N).

Exceptionally, the Pyu dsDNA solutions with ds3C3G or dsCTTAAG showed a significant increase in infrared absorption bands by CEDS using a corresponding sequence compared to the other Pyu oligo-dsDNA solution observed in this study. Meanwhile, Puy dsDNA solutions with ds3G3C or dsGAATTC showed a slight decrease in infrared absorption band by CEDS using a corresponding sequence and a significant decrease by CEDS using a nonspecific sequence, 12A, compared to the untreated oligo-dsDNA solutions (Fig. 17 D and L).

The solution containing the oligo-dsCACGTG (three Pyu dsDNA segments of the canonical E-box sequence) showed a significant increase in infrared absorption band by CACGTG*-CEDS and a slight increase by 12A*-CEDS (Fig. 17 G). In contrast, the oligo-dsTTAGGG (a Pyu dsDNA of telomere repeat sequence) showed only a slight increase in infrared absorption band by TTAGGG*-CEDS and a significant decrease by 12A*-CEDS (Fig. 17 H).

The data suggest that CEDS using a corresponding sequence can significantly affect the N-H stretching of hydrogen bonds in a manner specific to the target dsDNA sequence compared to CEDS using a nonspecific sequence, 12A, and the untreated control.

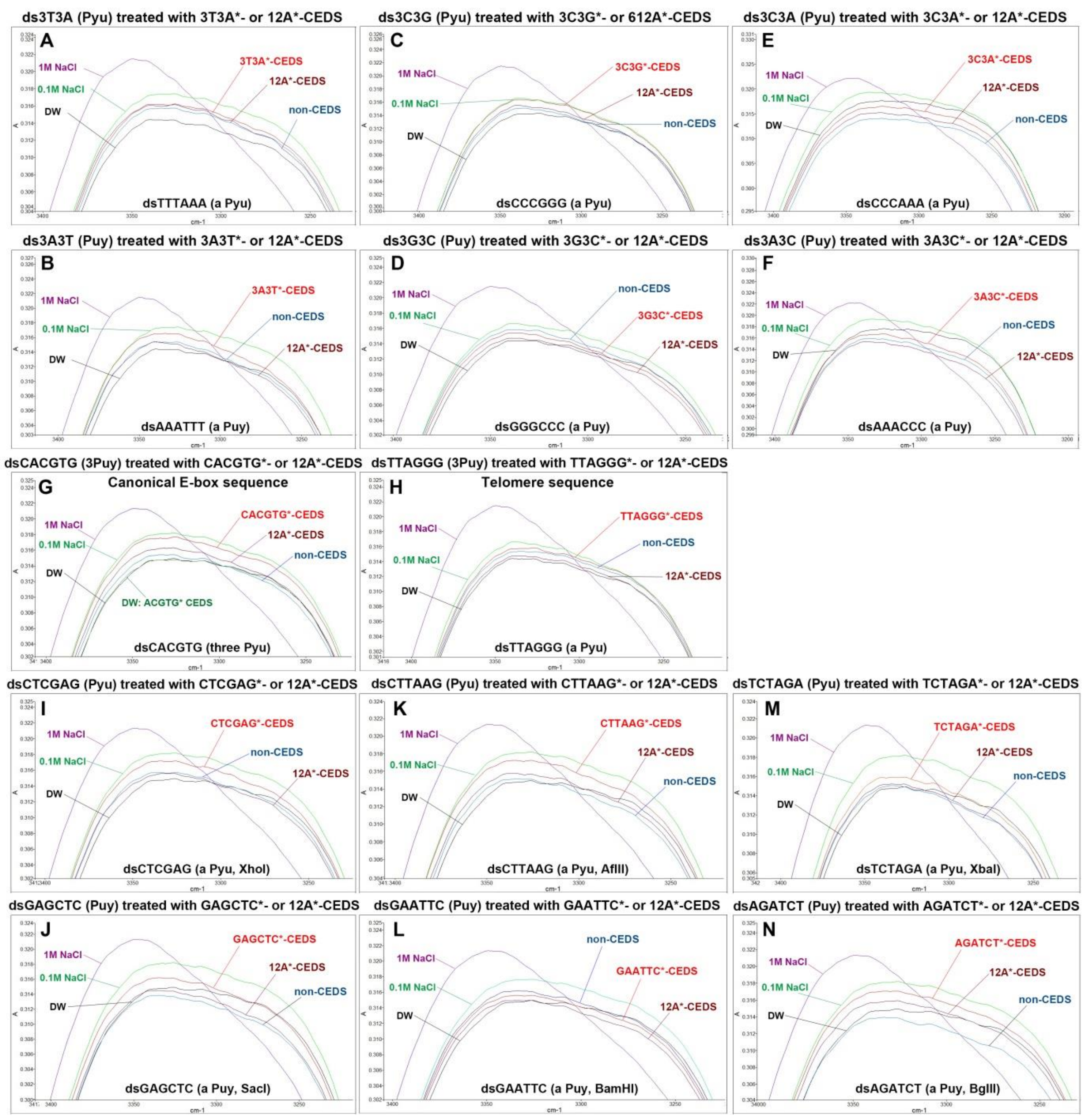

**Figure 17.** FT-IR analysis for CEDS effect on different oligo-dsDNAs. The six sequence-long dsDNAs including the simple and complex oligo-dsDNAs composed of Pyu or Puy dsDNA segment(s) showed a trend of increased IR-absorption at about 3700-2800 $cm^{-1}$ band centered at 3350-3300 $cm^{-1}$ by corresponding sequence*-CEDS compared to nonspecific sequence, 12A*-CEDS and the untreated control.

**CEDS effect on the binding of ethidium bromide (EtBr) and spermidine in oligo-dsDNAs**

***1) CEDS effect on the EtBr DNA intercalation***

Since EtBr intercalates with the base pairs of dsDNA, causing significant conformational changes that may interfere with the essential functions of dsDNA[10], it would be informative to know CEDS effect on EtBr-intercalated dsDNA *in vitro*. When dsDNA was intercalated by EtBr, it became stiffened, thereby their conformational changes could be detected by UV and fluorescence spectroscopy. In this study, we prepared ds4(3T3A), ds6(2C2A), and ds6C6A solutions (0.1M NaCl) at 100 pmole/µL as the same method described above, and mixed the different oligo-dsDNAs with varying concentrations of EtBr (1-50 µg/mL) and performed CEDS with the corresponding sequences separately. The HPLC method was used to analyze the DNA conformational changes of each oligo-dsDNA.

The ds4(3T3A) solution was mixed with EtBr at 1 and 4 µg/mL at room temperature for 10 min separately, and treated with 4(3T3A)*-CEDS for 180 min. During CEDS, 50 µL of ds4(3T3A) solution was obtained at 0, 30, 60, 90, 120, 150, 180 min, and immediately each sample was analyzed using HPLC with a reverse phase silica column and a running buffer of 0.1M NaCl solution at a flow rate of 0.3 mL/min.

When the ds4(3T3A) solution was separately mixed with EtBr at 1 and 4 µg/mL for 10 min, HPLC peak of ds4(3T3A) decreased up to 77.9% and 28.3% of the untreated control level, and recovered by 4(3T3A)*-CEDS up to 98.5% and 82.2% of the untreated control level until 180 min, respectively (Fig. 18 A). The data show that the conformation of ds4(3T3A) was condensed by EtBr intercalation, depending on the concentration of EtBr, and that the EtBr-induced DNA condensation was partly recovered by 4(3T3A)*-CEDS treatment within 180 min.

To know the EtBr-dependent dsDNA condensation, the ds6(2C2A) solution was mixed with EtBr at 3, 10, and 50 µg/mL by adding concentrated EtBr solution (1 mg/mL) at room temperature for 10 min separately. And then each sample was treated with 6(2C2A)*-CEDS for 150 min. Each 50 µL of ds6(2C2A) solution was obtained after EtBr mixing at 3, 10, and 50 µg/mL, and during 6(2C2A)*-CEDS at 30, 60, 90, 120, and 150 min. The samples were analyzed using HPLC with a reverse-phase silica column and a running buffer of 0.1M NaCl solution at a flow rate of 0.3 mL/min.

The HPLC peak area of ds6(2C2A) decreased up to 98.6%, 91.8%, and 78.3% of the untreated control level with increasing concentrations of 3, 10, and 50 μg/mL EtBr. The maximum decrease induced by 50 μg/mL EtBr was gradually recovered up to 34.3% by 6(2C2A)*-CEDS after 150 min (Fig. 18 B). These data demonstrate that the HPLC peak area of oligo-dsDNA, ds6(2C2A) decreased depending on the concentration of EtBr, and can be partially recovered by 6(2C2A)*-CEDS depending on CEDS time.

To determine the specific removal of EtBr from oligo-dsDNA by CEDS, ds6C6A solution was mixed with EtBr at 10 μg/mL for 10 min. The mixture was separately treated with CEDS using a target sequence, 6C6A, or a random sequence, 3(ACGT), at room temperature for 180 min. After CEDS, 50 μL of ds6C6A solution was collected at 0, 30, 60, 90, 120, 150, and 180 min. The solution was then immediately analyzed using HPLC equipped with both UV spectroscopy and fluoroscopy. A reverse phase silica column and a running buffer of 0.1M NaCl solution at a flow rate of 0.3 mL/min were used.

In UV spectroscopic HPLC analysis, the ds6C6A solution containing 10 μg/mL EtBr showed a significant decrease in HPLC peak area up to 88.6% of the untreated control level. This change was significantly reversed by 6C6A*-CEDS, up to 28.6% after 180 min, while weakly reversed by random sequence 3(ACGT)*-CEDS, up to 4.3% (Figure 18G). HPLC fluoroscopy analysis detected the presence of EtBr in ds6C6A solution containing 10 μg/mL EtBr by excitation at 301 nm and emission at 603 nm, and EtBr was significantly removed from ds6C6A up to 6.6% of the untreated control level by 6C6A*-CEDS after 180 min, while slightly removed up to 2% by 3(ACGT)*-CEDS (Figure 18 C-F and H).

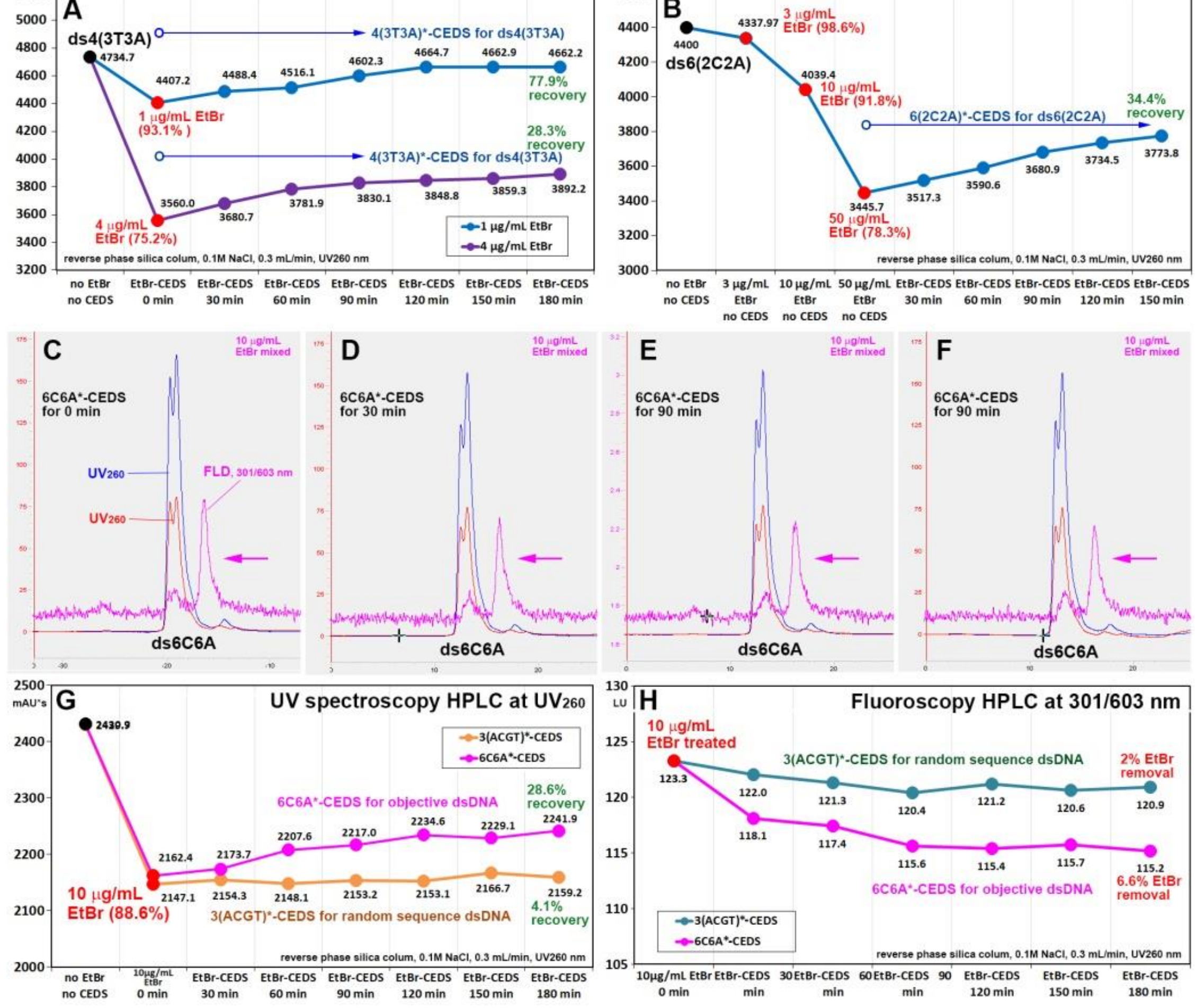

**Figure 18.** CEDS effect on the EtBr intercalation with oligo-dsDNAs through UV and fluorescence HPLC. A and B: ds4(2T2A) and ds6(2C2a) in 0.1 M NaCl solution were severely condensed depending on the concentration of EtBr at 1-50 μg/mL, and the DNA peak of condensed oligo-dsDNAs were gradually recovered by 4(2T2A)*-CEDS and 6(2C2A)*-CEDS, respectively, depending on CEDS time until 180 min through UV-HPLC analysis. C-H: ds6C6A was severely reduced by 10 μg/mL EtBr, and the DNA peak of reduced oligo-ds6C6A was markedly recovered by 6C6A*-CEDS depending on CEDS time until 180 min by 28.6% through UV-HPLC analysis, but slightly recovered by random sequence 3(ACGT)*-CEDS by only 4.1%. During the recovery period, fluorescence HPLC showed significant decrease of EtBr fluorescence in the DNA peak of ds6C6A by 6C6A*-CEDS depending on CEDS time by 6.6%, but weak decrease by random sequence 3(ACGT)*-CEDS by only 2%.

***2) CEDS effect on the spermidine DNA condensation***

Spermidine readily binds to dsDNA, causing condensation and loss of DNA function. The aim of this study is to assess CEDS effect on spermidine-induced oligo-dsDNA condensation and determine if the condensed dsDNA can be recovered by CEDS. The ds3(4C4A) was dissolved in 0.1M NaCl solution at 100 pmole/μL and mixed with spermidine at varying concentrations of 1, 2, 4, 8, 12, 16, 20, 30, 40, 50, and 100 mM separately. Each sample was treated with 3(4C4A)*-CEDS for 30 min and then analyzed using HPLC with a reverse phase silica column and running buffer of 0.1M NaCl solution at a flow rate of 0.3 mL/min. The HPLC peak areas of ds3(4C4A) treated with CEDS were compared to those of the untreated controls.

The samples mixed with spermidine at various concentrations of 1, 2, 4, 8, 12, 16, 20, 30, 40, 50, and 100 mM showed significant decreases in HPLC peak area depending on the concentration of spermidine, resulting in a steep downward slope up to 77% of the untreated control level, whereas when the samples mixed with spermidine at different concentrations of 1, 2, 4, 8, 12, 16, 20, 30, 40, 50, and 100 mM were separately treated with 3(4C4A)*-CEDS, their HPLC peak area showed a relatively slow decrease with a gentle downward slope up to 86. 2% of the untreated control level compared to the untreated control (Figure 19 A-C).

These results indicate that the oligo-dsDNA, ds3(4C4A), was significantly condensed by spermidine depending on the concentration of spermidine (1-100 mM) in the HPLC analysis, and the spermidine-induced condensation of ds3(4C4A) was slightly decreased by 3(4C4A)*-CEDS compared to the untreated controls as the spermidine concentration increased, indicating the partial removal of spermidine from ds3(4C4A) by 3(4C4A)*-CEDS.

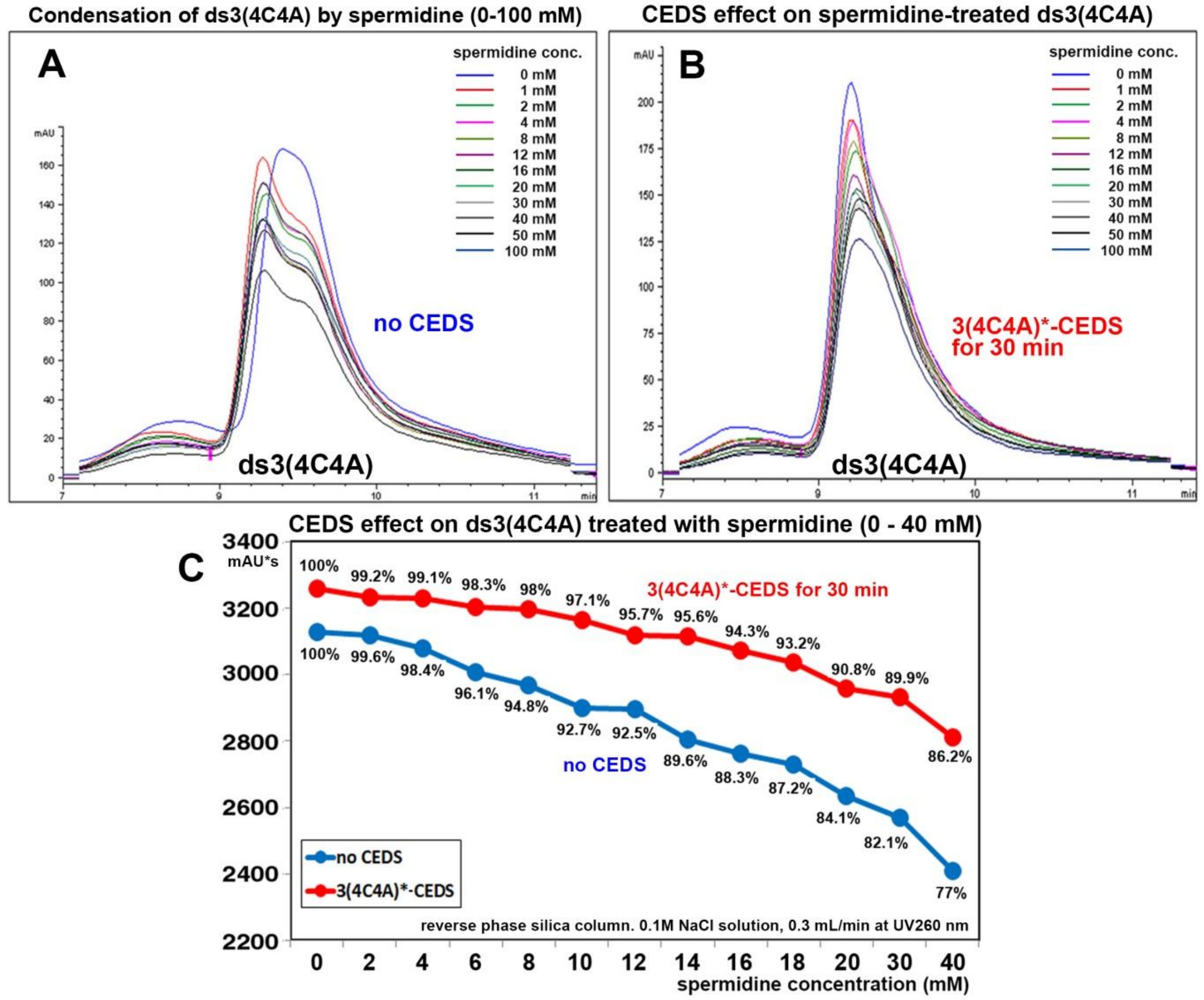

**Figure 19.** CEDS effect on the spermidine-induced DNA condensation of oligo-dsDNA through HPLC analysis. The ds3(4C4A) in 0.1 M NaCl solution was mixed with different concentration of spermidine, 1-100 mM, and resulted in severe decrease of DNA peak in HPLC depending on the concentration of spermidine (A). The decrease of DNA peak by spermidine-induced DNA condensation was much alleviated by 3(4C4A)*-CEDS for 30 min (B). 3(4C4A)*-CEDS effect on spermidine-induced DNA condensation at 0-40 mM was plotted into a line graph (C).

To know the recovery after spermidine-induced DNA condensation in a sequence-specific manner, the ds2(6C6A) was dissolved in 0.1M NaCl solution at 100 pmole/μL and mixed with spermidine at 5 mM for 10 min, and then treated with CEDS using a target sequence, 2(6C6A), or the random sequence, 3(ACGT), at room temperature for 180 min. During CEDS, 50 μL of ds2(6C6A) solution was obtained at 0, 30, 60, 90, 120, 150, and 180 min. Each sample was immediately analyzed using HPLC with a reverse phase silica column and a running buffer of 0.1M NaCl solution at a flow rate of 0.3 mL/min. The changes in DNA peak area induced by the 2(6C6A)*CEDS were compared with those induced by the 3(ACGT)*CEDS and plotted on a graph.

The ds2(6C6A) showed a marked reduction of the HPLC peak area up to 88.1% of the untreated control level by mixing with spermidine at 5 mM concentration in the absence of $Ca^{++}$, and the reduced HPLC peak area was gradually recovered by CEDS using a targeted sequence, 2(6C6A)*-CEDS, up to 28.9% of the spermidine (5 mM)-induced decrease after 180 min, whereas it was rather decreased by CEDS using a random sequence, 3(ACGT)*-CEDS, up to 46.7% (Fig. 20 A-C).

The data indicate that spermidine significantly condensed the oligo-dsDNA, ds2(6C6A), and this effect was reversed by CEDS using a target sequence, 2(6C6A)*-CEDS, but not by CEDS using a random sequence, 3(ACGT)*-CEDS. Therefore, it is proposed that CEDS can renature the oligo-dsDNA, ds2(6C6A), to partially recover from the spermidine-induced deformation of DNA conformation by the HBMR in a sequence-specific manner.

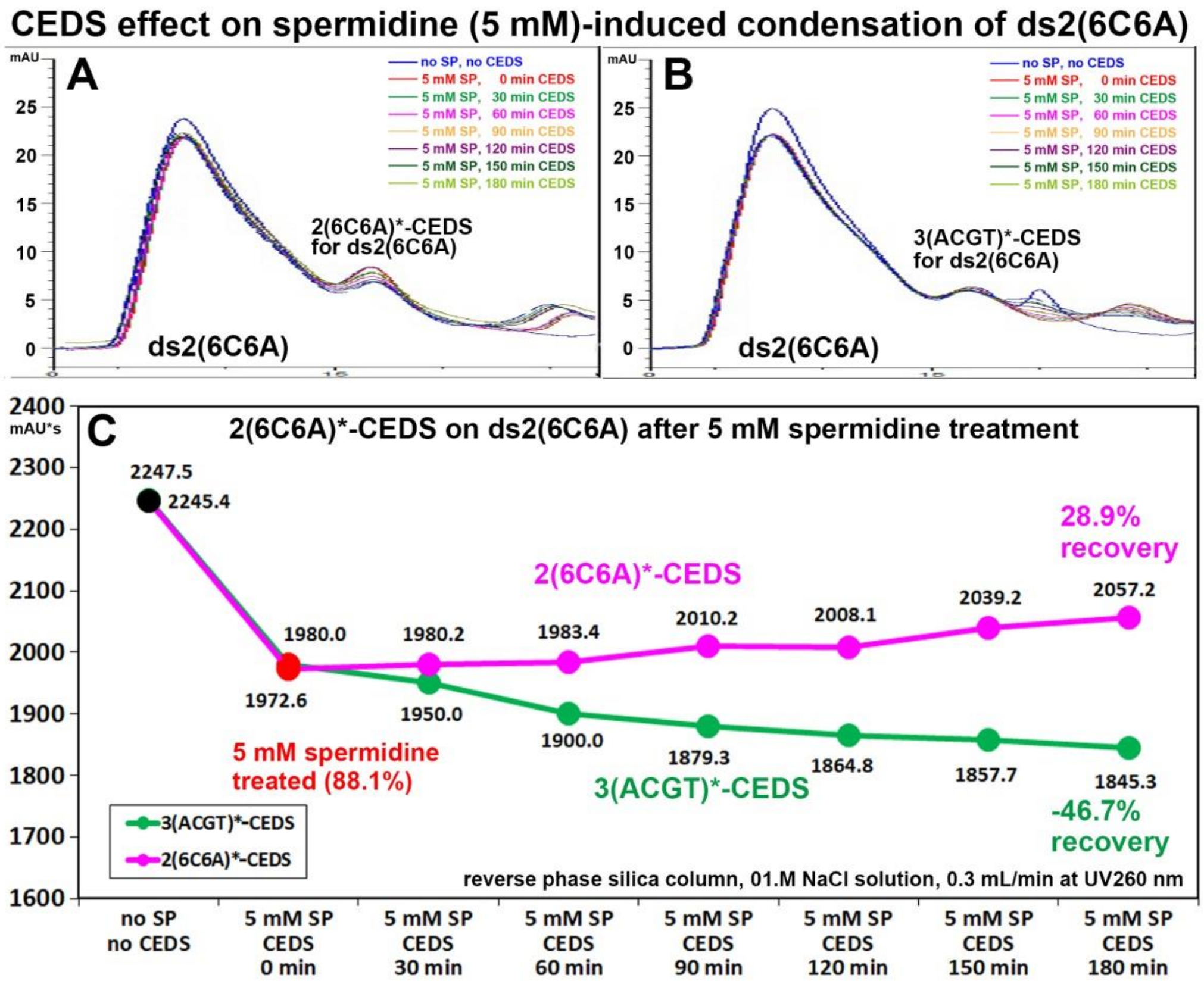

**Figure 20.** HPLC analysis for the comparison of CEDS effect on the oligo-dsDNA condensed by spermidine (SP) at 5 mM between by using a target sequence and a random sequence*-CEDSs (A and B). ds(6C6A) in 0.1M NaCl solution showed severe decrease of DNA peak area by 5 mM spermidine, and the reduced DNA peak area was gradually recovered by 2(6C6A)*-CEDS depending on CEDS time until 180 min by 28.9% of spermidine (5 mM)-induced decrease, while rather decreased by random sequence 3(ACGT)*-CEDS by 46.7% (C).

**CEDS effect on restriction endonuclease (RE) digestion of plasmid DNA**

Plasmid DNA (pBluescript II SK(-) Y166 clone) was digested with different restriction enzymes, including BamHI, EcoRI, KpnI, NotI, PstI, SacI, XhaI, and XhoI, under CEDS using a corresponding restriction site sequence separately. For instance, 5μg of Y166 plasmid DNA was digested with 5U of XhoI enzyme under CEDS using a XhoI site sequence (CTCGAG) in the incubator at 37℃ for 30 min. The DNA product underwent electrophoresis on 1% agarose gel, stained with EtBr (0.1μg/mL) for 1 minute, and was detected under UV illumination. To prevent nonspecific overstaining of EtBr, the DNA product was pre-stained with EtBr (1 μg/mL) for 1 minute. It was then electrophoresed on 1% agarose gel and detected under UV illumination. The negative and positive controls were performed simultaneously without CEDS and with CEDS using a random sequence, (2(ACGT)*-CEDS, respectively.

The unique DNA band found in the electrophoresis may indicate the amount of linear DNA digested by the RE enzyme compared to the smeared multiple bands of the original plasmid DNAs. Each plasmid DNA digested by BamHI, EcoRI, KpnI, NotI, PstI, SacI, XhaI and XhoI RE enzymes under CEDS using a corresponding RE binding site sequence, i.e. GGATCC*-CEDS, GAATTC*-CEDS, GGTACC*-CEDS, GCGGCCGC*-CEDS, CTGCAG*-CEDS, GAGCTC*-CEDS, TCTAGA*-CEDS and CTCGAG*-CEDS, respectively, showed the stronger DNA bands compared to the negative and positive controls (Fig. 21 A and C-I). The DNA products of BamHI and XhoI digestion under GGATCC* and CTCGAG*-CEDS, respectively, were examined by pre-electrophoresis EtBr staining and then clearly showed the stronger DNA bands compared to the negative and positive controls (Fig. 21 B and J).

The data suggest that the RE digestion of plasmid DNA is enhanced by CEDS using a corresponding RE site sequence, indicating CEDS facilitate the RE site dsDNA to be readily digested by providing the unique DNA conformation for the binding of a specific restriction enzyme.

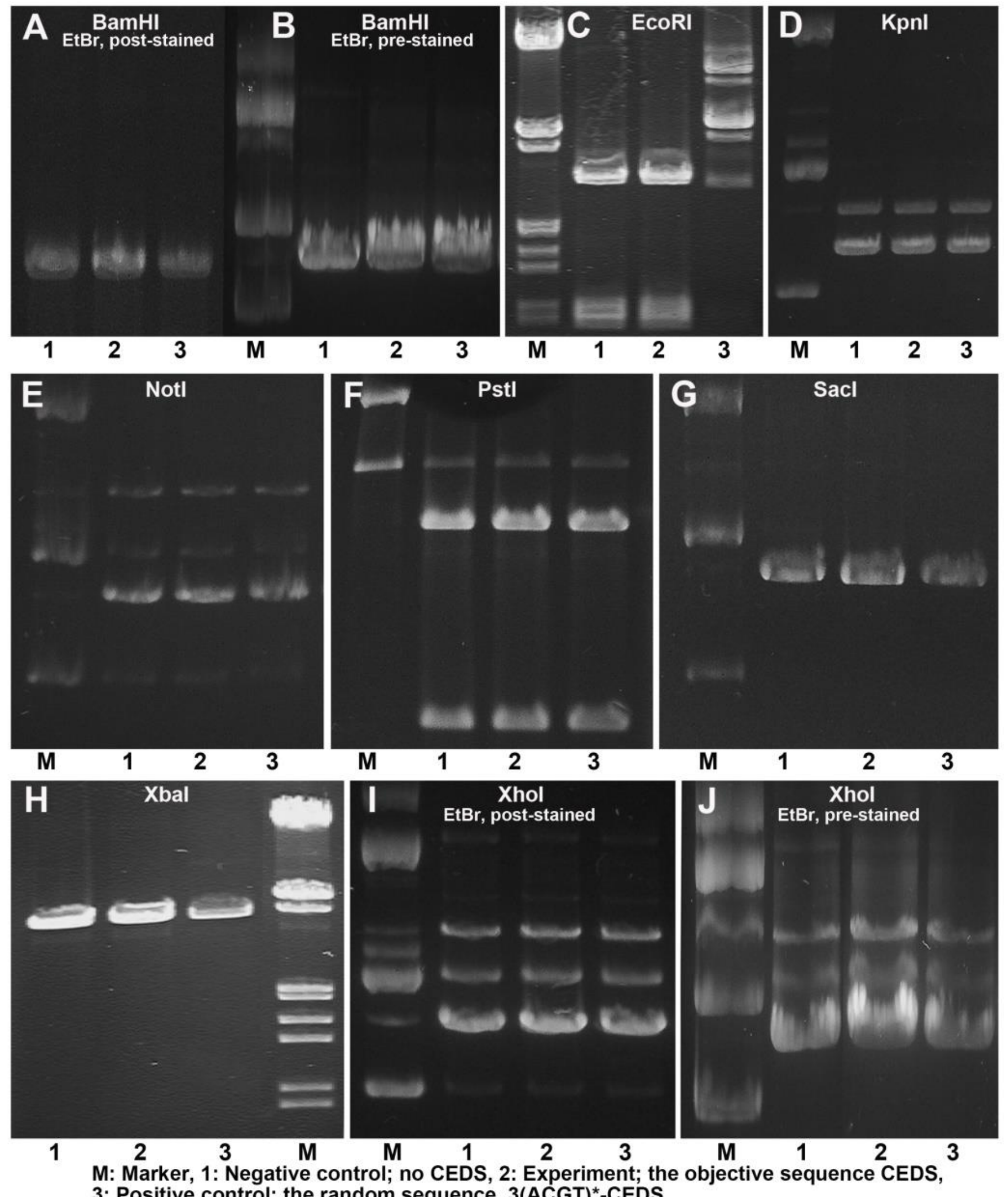

**Figure 21.** CEDS effect on different RE digestions of plasmid DNA (pBluescript II SK(-), Y166 clone). BamHI (A and B), EcoRI (C), KpnI (D), NotI (E), PstI (F), SacI (G), XhaI (H), and XhoI (I and J) digestion under CEDS using a corresponding RE site sequence, GGATCC, GAATTC, GGTACC, GCGGCCGC, CTGCAG, GAGCTC, TCTAGA, and CTCGAG, respectively, produced the stronger DNA bands compared to the negative and positive controls. In the pre-electrophoresis EtBr staining, the DNA bands digested by BamHI and XhoI, under GGATCC* and CTCGAG*-CEDS, respectively, appeared stronger than those of the negative and positive controls. M: Marker, 1: Negative control with no CEDS, 2: Experiment; the target sequence*-CEDS, 3: Positive control; random sequence, 3(ACGT)*-CEDS

To evaluate CEDS effect on the RE digestion of plasmid DNA depending on CEDS time, each 30 μg of plasmid DNA was digested with BamHI, EcoRI, KpnI, PstI, SacI, XhaI, and XhoI, under CEDS using a corresponding restriction site sequence, GGATCC, GAATTC, GGTACC, CTGCAG, GAGCTC, TCTAGA, and CTCGAG, respectively, in the incubator at 37°C for 100 min. Additionally, the RE digestion without CEDS and with random sequence (2(ACGT))*-CEDS were performed simultaneously under the same conditions as the above experiments.

During the RE digestion under CEDS, each 2 μg DNA sample incubated for 1, 10, 20, 30, 40, 50, 60, 70, 80, 90 and 100 min was analyzed by HPLC using a non-hydrophobic silica column, 0.1M NaCl solution mobile buffer (0.5 mL/min) at UV260. The increase of the HPLC peak may indicate the conformational changes from the supercoiled plasmid DNAs to the linear DNAs. The incremental data as a function of RE digestion were plotted and compared.

The HPLC peak area (mAU*s) represents both the reduction of the original plasmid DNA and the increase of the linear RE digestion product at the same time, which increased proportionally during the RE digestion time (Fig. 22 A). CEDS using a corresponding RE site sequence rapidly increased the digestion of plasmid DNA with BamHI, EcoRI, KpnI, PstI, SacI, XbaI, or XhoI restriction enzymes during 10-60 min of CEDS time and plateaued at 100 min compared to the negative and positive controls (Fig. 22 B-H).

The data can show the RE digestion of plasmid DNA rapidly increased by CEDS using a corresponding restriction site sequence during 10-60 min of CEDS time, and CEDS-induced increase of each RE digestion was significantly higher than those by no CEDS and CEDS using random sequence (2(ACGT)) up to 100 min of CEDS time. .

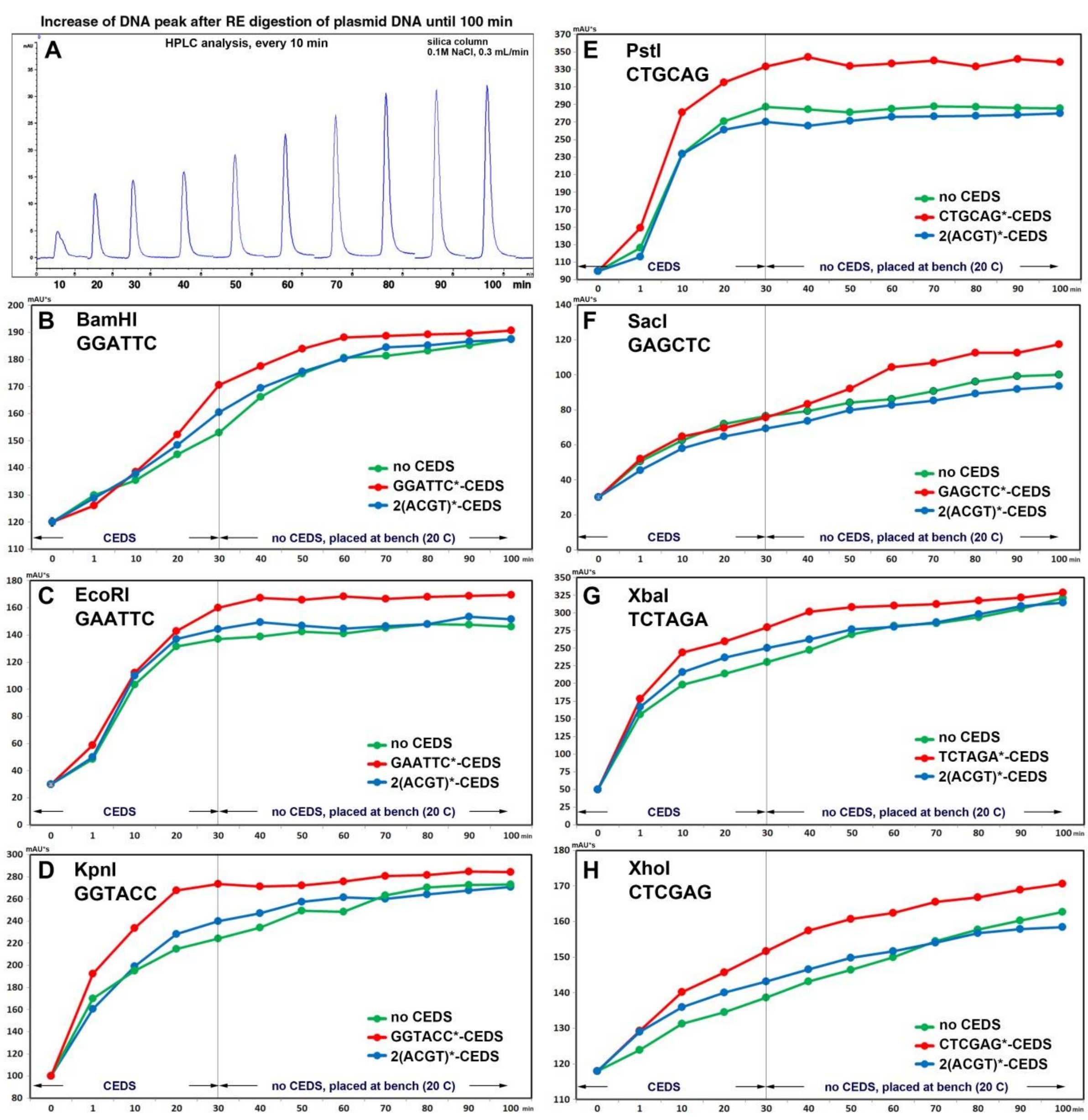

**Figure 22.** HPLC analysis for different RE digestions of plasmid DNA (Y166) under the corresponding restriction site sequence for 100 min. A: The linear DNA peaks by BamHI digestion with no CEDS. B-H: Line graphs comparing CEDS using a target sequence, the random sequence (2(ACGT)), and the untreated control. B: BamHI digestion under GGATCC*-CEDS. C: EcoRI digestion under GAATTC*-CEDS. D: KpnI digestion under GGTACC*-CEDS, E: PstI digestion under CTGCAG*-CEDS. F: SacI digestion under GAGCTC*-CEDS, G: XbaI digestion under TCTAGA*-CEDS. H: XhoI digestion under CTCGAG*-CEDS. Red lines; experimental groups. Green lines; negative controls without CEDS. Blue lines; positive controls with random sequence, 2(ACGT)*-CEDS.

**CEDS effect on *in vitro* RNA transcription from plasmid DNA (pBluescript SK(-))**

The *in vitro* RNA transcription assay was performed using the linear plasmid DNAs (pBluescript SK(-)) inserted with human VEGFA, vWF, or elafin cDNAs (500-600 bps) separately. Each template DNA (approximately 0.8 μg) was added to 20 μL of transcription buffer (40mM Tris-HCl, pH 8) for the *in vitro* transcription reaction. The reaction mixture (0, 10 mM MgCl2, 10 mM DTT, 4 mM spermidine, 10 mM NaCl, 50 μg/mL BSA) was prepared with 1 unit/μL RNAse inhibitor, 0.4 units/μL RNA polymerase and 0.5 mM NTP mixture (USB Corp. Ohio, USA). Meanwhile, each template DNA of VEGFA, vWF, or elafin was placed under CEDS using a T7, SP6, or T3, promoter sequence, i.e., the TAATACGACTCACTATAGGG-CEDS, ATTTAGGTGACACTATA-CEDS, or AATTAACCCTCACTAAAGGG-CEDS, respectively, in the incubator at 37°C for 20 or 40 min. After the experiment, the mixtures were treated with 1 unit of DNase I and incubated for 10 min at 37°C to degrade the template DNA. The RNA products were then immediately electrophoresed using a 1.5% formaldehyde agarose gel (20 mM MOPS, 6% formaldehyde) and DEPC-based buffer, stained with EtBr, and detected under UV illumination.

The study found that *in vitro* RNA transcription from human VEGF, vWF, and elafin cDNAs subcloned in plasmid DNA was increased up to 8.9%, 12.9%, and 33.4% after 20 min and up to 6.2%, 2.4%, and 2.4% after 40 min, respectively, by CEDS using a corresponding promoter (T7, SP5, or T3) sequence (Fig. 23 A-F). These results suggest that CEDS enhances *in vitro* RNA transcription by stimulating the target promoter sequence to recruit and activate RNA polymerase.

In particular, when vWF RNA transcription was performed under CEDS using either a forward SP6 promoter sequence or a reverse SP6 promoter sequence for 20 min, CEDS using a forward SP6 promoter sequence produced up to 21.1% more RNA transcription after 20 min than the untreated control, whereas CEDS using a reverse SP6 promoter sequence produced only a 9% increase in RNA transcription (Fig. 23 G and H). The results suggest that CEDS using a forward SP6 promoter sequence could strongly activate RNA transcription, while CEDS using a reverse SP6 promoter sequence could weakly activate.

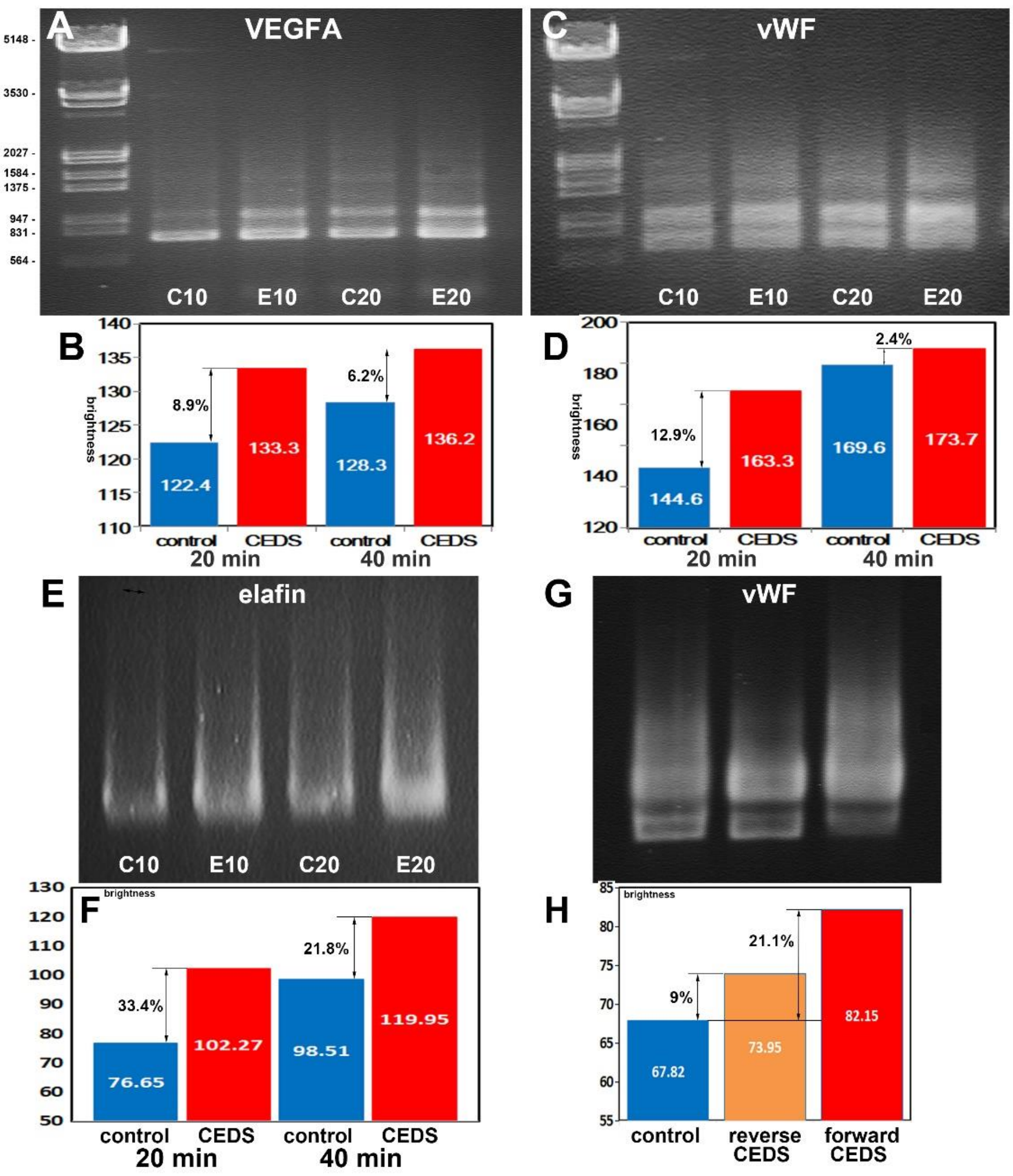

**Figure 23.** *In vitro* RNA transcription of human VEGFA (A and B), vWF (C and D), elafin (E and F) mRNAs from plasmid cDNAs under CEDS using a T7, SP6, and T3 promoter sequence, respectively, and analysis by densitometer. *In vitro* RNA transcription of human vWF cDNA under forward or reverse SP6 promoter sequence-CEDS (G and H). CEDS using a corresponding promoter sequence resulted in a higher amount of RNA transcript of human VEGFA, vWF, elafin cDNAs than the untreated control, and CEDS using a forward SP6 promoter sequence resulted in a higher amount of RNA transcript of human vWF cDNA than CEDS using a reverse SP6 promoter sequence.

To determine if CEDS enhances *in vitro* RNA transcription depending on CEDS time, 10 μg of pBluescript SK (-) vector containing human elafin cDNA in its multi-cloning site was used for *in vitro* RNA transcription to obtain 200 μL reaction mixture as described above. Subsequently, each 100 μL of the reaction mixture was treated with CEDS using a forward T3 promoter sequence or the reverse T3 promoter sequence in an incubator at 37°C for 120 min. During the incubation under CEDS, 10 μL of the reaction mixture was taken at 30, 60, 90, and 120 min. Each sample was treated with 1 unit of DNase I and incubated for 10 min at 37℃ to degrade the template DNA. The RNA products were then immediately electrophoresed using a 1.5% formaldehyde agarose gel (20 mM MOPS, 6% formaldehyde) and DEPC-based buffer. Finally, the RNA products were stained with EtBr and detected under UV illumination. The brightness of the RNA band was measured with a densitometer and plotted on a graph.

CEDS using a forward T3 promoter sequence increased the *in vitro* RNA transcription of the human elafin mRNA from plasmid vector by 11.9% at 30 min, 23.3% at 60 min, 11.2% at 90 min, and 21.3% at 120 min (Fig. 24 A and B), whereas CEDS using a reverse T3 promoter sequence decreased the *in vitro* RNA transcription by 2.3% at 30 min, 0.3% at 60 min, 1.6% at 90 min, and 3.4% at 120 min (Fig. 24 C and D). The data indicate that CEDS using an objective DNA sequence, the forward T3 promoter sequence, consistently enahanced RNA transcription for 120 min, while CEDS using an inhibitory DNA sequence, the reverse T3 promoter sequence, slightly inhibited RNA transcription.

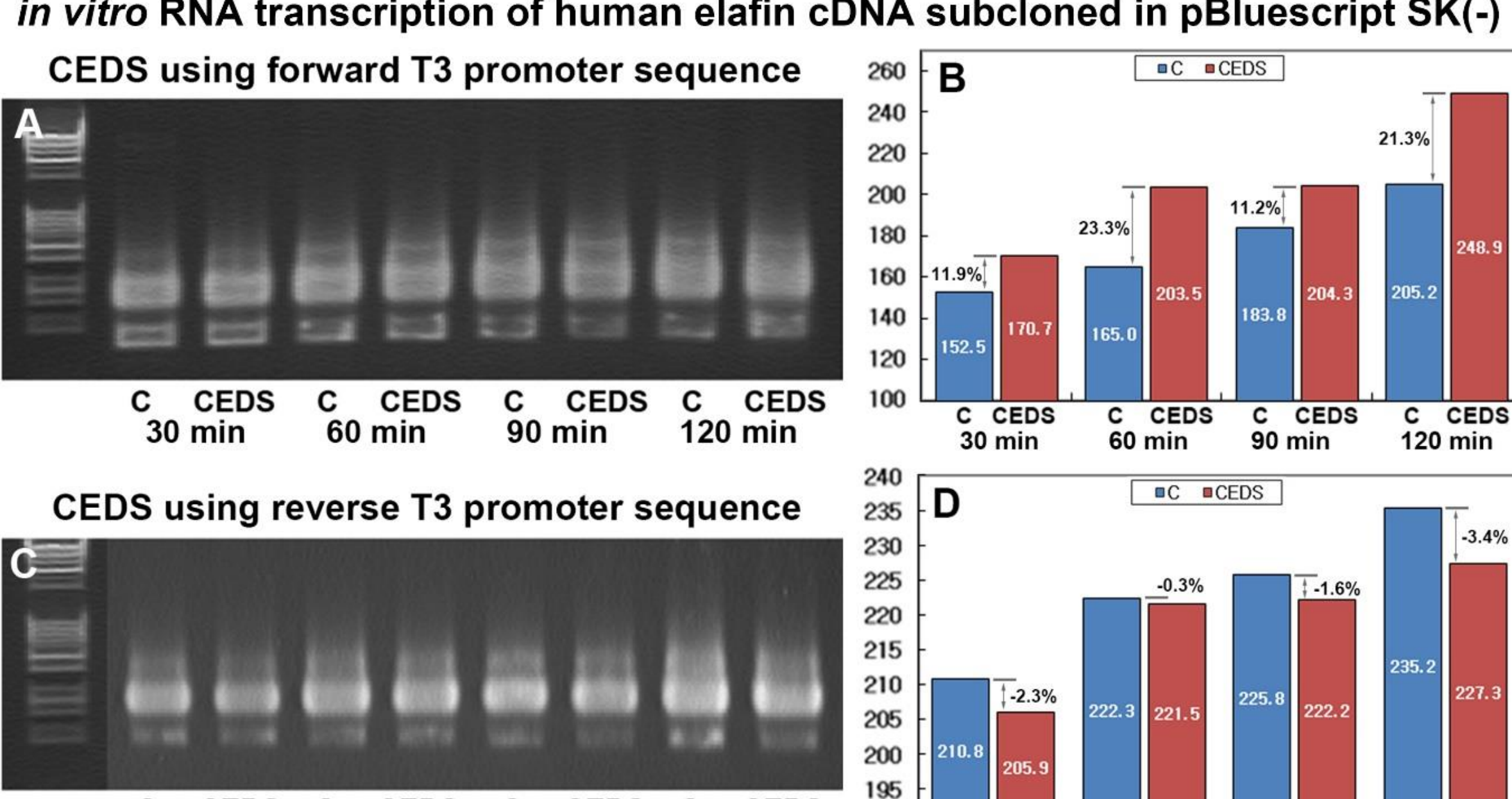

**Figure 24.** *In vitro* RNA transcription of human elafin mRNA from plasmid DNA under CEDS using a forward or reverse T3 promoter sequence. CEDS using a forward T3 promoter sequence (A and B) resulted in a higher amount of RNA transcript of human elafin than CEDS using a reverse T3 promoter sequence (C and D) consistently throughout CEDS time.

**CEDS effect on reporter protein production of plasmid DNA vectors**

***Production of green fluorescent protein (GFP) from pE-GFP-1 by CEDS***

E. coli cells were transfected with the pE-GFP-1 vector (Clontech, USA) and cultivated in Luria Bertani (LB) medium supplemented with kanamycin (50 μg/mL). The pE-GFP-1 vector contains a green fluorescent protein (GFP) gene, inserted at the 5' flanking end of the T3 promoter. The E. coli cell culture was prepared at a standard concentration of 0.5 $OD_{600}$, as measured by UV spectrometer. The standard E. coli cell culture was incubated under the T3 promoter sequence-CEDS at 20-25 Gauss, 37℃ for 30 min. The positive and negative control groups were also incubated with 1.5 mL of standard E. coli broth under CEDS using a random sequence (6(ACGT))-CEDS and no CEDS, respectively. After CEDS treatment, the E. coli culture was incubated in a shaking incubator at 37°C for 1 hour. The supernatant was obtained after centrifugation at 1000x g, and GFP was detected using a spectrofluorometer (FP-6500, Jasco, Japan) with 488 nm excitation and 507 nm emission.

The experimental group subjected to CEDS treatment using a T3 promoter sequence to target the GFP gene in the pE-GFP-1 vector consistently generated more GFP than the positive and negative control groups for up to 5 hours during the experiment (Fig. 25 A-C). When the culture of pE-GFP-1 vector-transfected E. coli was treated with CEDS using a T3 promoter sequence with low magnetic field, 10 Gauss, for 3 and 5 hours, the culture products showed stronger GFP fluorescence on nitrocellulose membrane by 20% and 16.7%, respectively, than the untreated control (Fig. 25A-C).

Specifically, the GFP production under the T3 promoter sequence-CEDS showed a significant increase of approximately 70.6% and 66.7% at 3 and 5 hours, respectively, compared to the negative control performed without CEDS, while the GFP production under random sequence-CEDS, 3(ACGT)-CEDS, showed an increase of only 33.3% and 6.1% at 3 and 5 hours of the culture, respectively (Fig. 25 D).

The results suggest the steady increase in the GFP levels after CEDS treatment throughout the experimental phase could imply the activation of target DNA by T3 promoter sequence-CEDS. In contrast to the aforementioned CEDS experiments that employed synthetic oligo-dsDNAs or *in vitro* plasmid DNAs, the GFP expression experiment utilized the supercoiled plasmid DNAs in E. coli. This *in vivo* experiment thus indicates that CEDS can affect the functioning of even native genes within cells.

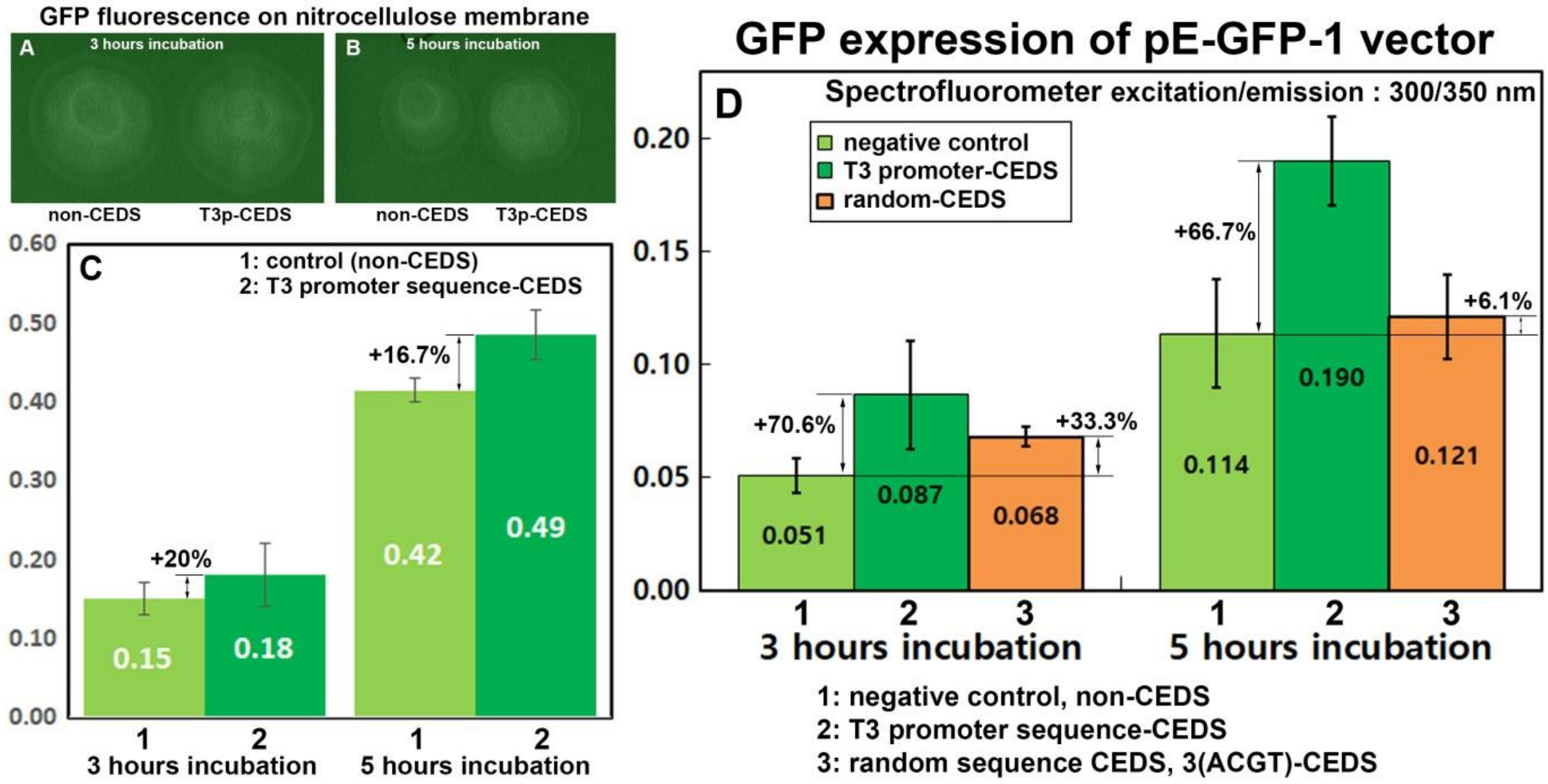

**Figure 25.** E. coli culture for green fluorescent protein (GFP) expression under CEDS using a T3 promoter sequence. A-C: The comparison between T3 promoter sequence-CEDS and non-CEDS on the culture of pE-GFP-1 vector-transfected E. coli. The 3 and 5 hours culture showed stronger GFP fluorescence on nitrocellulose membrane by 20% and 16.7%, respectively, than the untreated control. The densitometer data of A and B were plotted into a graph (C). D: The experimental group, which was treated with CEDS using a T3 promoter sequence, showed a higher fluorescence compared to the positive control groups treated with CEDS using a random sequence, 3(ACGT)-CEDS, as well as the negative control group that was conducted without CEDS. Proportionate data from multiple repeated experiments revealed significant differences among the groups ($p<0.028$).

***Production of β-galactosidase from pBluescript SK(-) vector by CEDS***

The pBluescript SK(-) vector from Stratagene (USA) contains a LacZ gene under the 5' and 3' flanking T3 and T7 promoters, respectively. The T3 promoter encodes the LacZ and ampicillin-resistance (ApR) genes in the forward direction, while the reverse direction does not. Therefore, the transcriptional activity by CEDS is detectable depending on the forward or reverse direction of the T3 promoter sequence through E. coli cultivation. The standard E. coli culture mixture was generated by introducing the vector into E. coli and cultivating it in LB media supplemented with 100 μg/mL ampicillin. The concentration of the mixture was 0.5 at UV600. For the experiment, 1.5 mL of the standard E. coli culture mix in 15 mL tube was treated with the forward T3 promoter sequence-CEDS at 37°C for 30 min. The positive control group was administered with either the forward T7 promoter sequence or the reverse T3 promoter sequence-CEDS. The negative control group was incubated without CEDS under the same conditions.

First of all, to know the *in vitro* response of oligo-ssDNAs or dsDNAs to the T3 promoter sequence-CEDS, the forward T3 promoter sequence (AATTAACCCTCACTAAAGGG) ssDNA and T3 promoter sequence dsDNA in 0.1M NaCl solution at 10 pmole/μL were prepared separately as previously described. The samples were then treated with CEDS using a forward or reverse T3 promoter sequence at room temperature for 30 min. The resulting DNA products were immediately examined with HPLC using spectroscopy at UV 260 nm to know their conformational changes as previously described.

The oligo-dsDNA of the T3 promoter sequence showed a significant decrease in UV absorption by the forward T3 promoter sequence-CEDS, up to 6.5% compared to the untreated control. There was only a rare response (0.3% increase) by the reverse T3 promoter sequence-CEDS. On the other hand, the oligo-ssDNA of the T3 promoter sequence showed a significant increase in UV absorption by the forward T3 promoter sequence-CEDS, up to 8.7% compared to the untreated control. There appeared only a weak increase of up to 2.7% by the reverse T3 promoter sequence-CEDS (Fig. 26 A).

The data indicate that the forward T3 promoter sequence-CEDS affects the conformation of not only oligo-dsDNA but also oligo-ssDNAs of T3 promoter sequence, constricting and expanding them, respectively. In contrast, the reverse T3 promoter sequence-CEDS showed minimal changes. However, it was proposed that the oligo-dsDNA of the T3 promoter sequence be constrained by CEDS to a state conducive to DNA hybridization and transcription. In contrast, the oligo-ssDNA appears to be aberrantly expanded by CEDS due to the absence of a complementary DNA strand.

The β-galactosidase assay is a popular gene regulation experiment that employs a LacZ gene-containing vector, making it both unique and simple to conduct. This study assesses the expression of β-galactosidase from the LacZ gene of the pBluescript vector when treating CEDS using a T3 or T7 promoter sequences. The amount of β-galactosidase production was measured by the X-Gal chemical reaction, and the resulting blue color was detected by UV-spectrometer at 380 nm.

The E. coli culture transfected with the pBluescript SK (-) vector was prepared as previously described. It was then treated with CEDS using a forward or reverse T3 promoter sequence, and a mutated forward T3 promoter sequence (mutation-1, A12T) at room temperature for 12 hours separately. The treated cultures were immediately examined by spectroscopy at UV 380 nm for the X-Gal reaction. A weak magnetic field, 2-3 Gauss, was utilized for the prolonged CEDS exposure on cells during this process. To perform the β-galactosidase assay, the culture treated with CEDS was mixed with 1 mM IPTG and 10 mM X-Gal, and the resulting mixture was subjected to X-Gal detection using a UV spectrometer at 380 nm. The data obtained was then analyzed statistically.

The forward T3 promoter sequence-CEDS was observed to increase β-galactosidase production in cells by up to 6.2% of UV380 absorption compared to the untreated control, while the reverse T3 promoter sequence-CEDS demonstrated a slight increase of up to 1.9% in β-galactosidase production in cells. In contrast, the mutated T3 promoter sequence (mutation-1, A12T)-CEDS exhibited a slight increase in β-galactosidase production in the cells up to 2.5% of UV380 absorption (Fig. 26 B). The data indicates that CEDS using a forward T3 promoter sequence enhances the transcription of the LacZ gene in a sequence-specific manner from the pBluescript SK (-) vector in the cells, in comparison to CEDS using a reverse T3 promoter sequence or a mutated T3 promoter sequence.

Activation of the T3 promoter of the pBluescript SK (-) vector can enhance the production of ampicillin resistance gene, which is critical for the survival of host cells, particularly E. coli, in ampicillin-supplemented media. To evaluate cell survival, a straightforward CEDS assay was conducted using various motif sequences of plasmid DNA. 1.5 mL culture of E. coli transfected with the pBluescript SK (-) vector, 0.5 at UV600, was prepared in a 15 mL tube as described previously. The culture was treated with CEDS using various promoter sequences, including T3 and T7 promoter sequence, 12A*, 3(TA)*, along with a mutated T3 promoter sequence at GG19,20CC (mutation-2) or G20C (mutation-3) separately. The treatment was carried out at room temperature for 1 or 2 days. To prevent any magnetic hazards during the prolonged exposure of CEDS, the magnetic field strength was limited to 2-3 Gauss. After treatment, the culture was immediately examined for surviving cells by spectroscopy at UV 600 nm.

As a result, CEDS using a T3 promoter sequence increased the UV600 absorbance up to 6.2% after 1 day and 4.2% after 2 days compared to the untreated control. In contrast, CEDS using a T7 promoter sequence, 12A*, or 3(TA)* decreased the UV600 absorbance by 8.9%, 11%, and 8.8% after 1 day and 1.6%, 1.55%, and 1.4% after 2 days, respectively. On the other hand, CEDS using a mutated T3 promoter sequence (mutation-2, GG19, 20CC) decreased the UV600 absorbance up to 8% after 1 day and 11.9% after 2 days, while CEDS using a mutated T3 promoter sequence (mutation-3, G20C) increased the UV600 absorbance up to 10.3% after 1 day and 23.8% after 2 days (Fig. 26 C).

The data indicate that CEDS using a T3 promoter sequence enhanced the transcription of the ampicillin resistance gene, allowing the bacteria to survive in ampicillin-supplemented media. In contrast, CEDS using a T7 promoter sequence, nonspecific sequences (12A), or 6(TA) slightly reduced the transcription. Moreover, CEDS utilizing the mutated T3 promoter sequence (mutation-2, GG19,20CC), which has been replaced with a novel Pyu dsDNA segment at the 3' end of the T3 promoter, has a negative impact on gene transcription. In contrast, CEDS using a mutated T3 promoter sequence (mutation-3, G20C), which has been mutated only at the end of the T3 promoter with minimal modification of the Pyu dsDNA segment structure, has been observed to positively affect gene transcription in a manner similar to that observed in CEDS using a wild-type T3 promoter sequence.

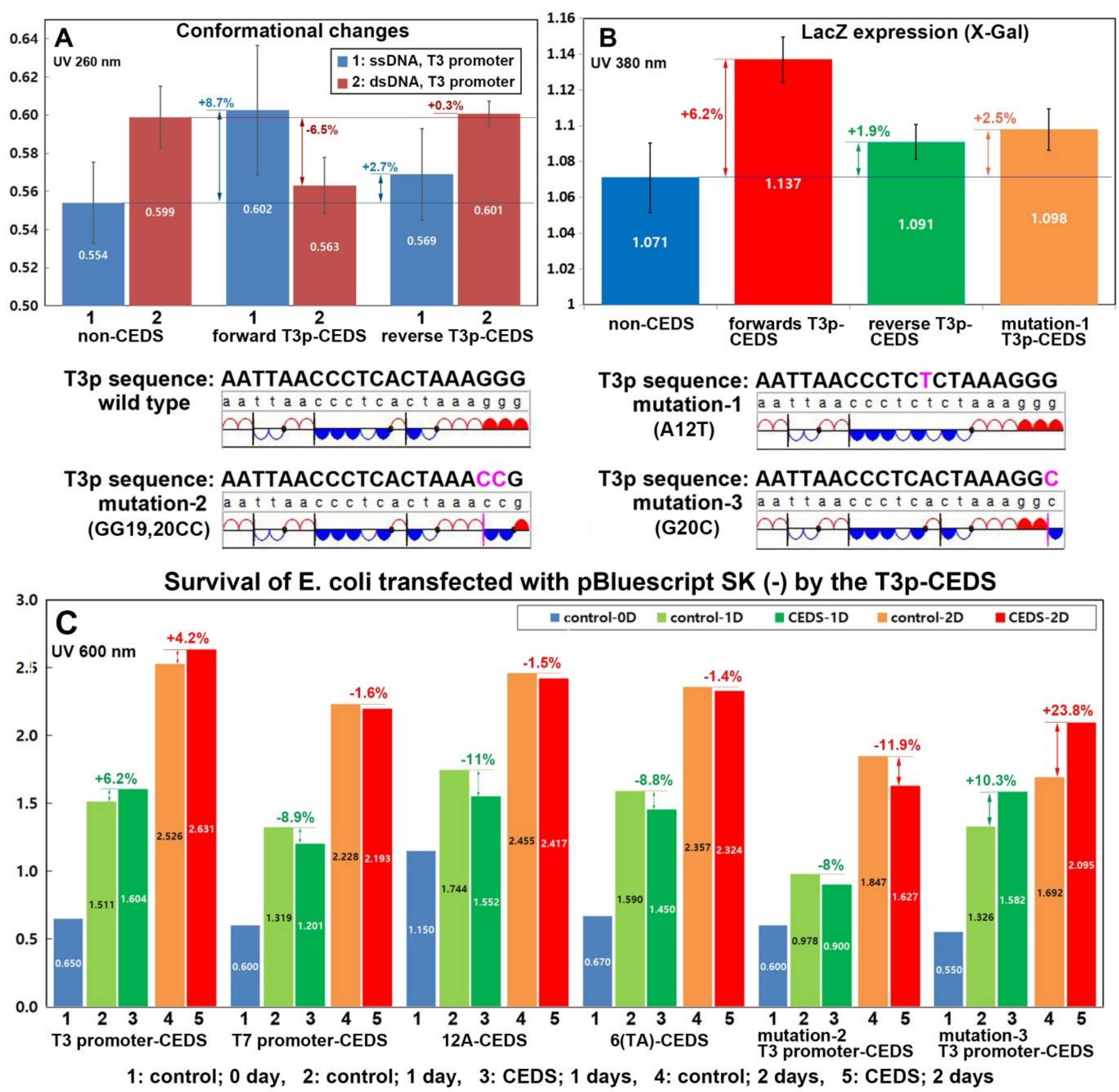

**Figure 26.** Spectroscopy analysis for the changes of target DNA conformation at UV 260 nm (A), LacZ expression at UV 380 nm (B), and E. coli survival in ampicillin-supplemented LB media at UV 600 nm (C). A: CEDS using a forward T3 promoter sequence differently affected the ssDNA and dsDNA of T3 promoter sequence, while CEDS using a reverse T3 promoter sequence showed only slight changes. B: E. coli transfected with pBluescript SK (-) showed a significant increase in X-Gal reaction by CEDS using a forward T3 promoter sequence. C: E. coli transfected with pBluescript SK (-) showed higher survival rates when treated with CEDS using a T3 promoter sequence after 1 and 2 days, compared to the untreated control.

The DNA double helix is composed of two single DNA strands that are complementary to each other. These strands are made up of four types of nucleotides: adenine (A), cytosine (C), guanine (G), and thymine (T). The nucleotides differ only in their base parts, which are either purines (A and G) or pyrimidines (C and T). This leads to the pairing of A with T and G with C, using two and three hydrogen bonds, respectively. To describe the molecular structures of dsDNA symmetrically, it could be defined that pyrimidines are negatively charged compared to purines, and that the G-C pair has stronger hydrogen bonding than the A-T pair due to the number of hydrogen bonds between base pairs. The polarities of DNA base pairs can be readily depicted by employing the distinctive symbols of each nucleotide in dsDNA (see Fig. 1).

The polarity of DNA base pairs arises from the dielectric properties of nucleotides and the dipolar hydrogen bonds between purine and pyrimidine in dsDNA. The polarity functions as a hybridization potential between two complementary base pairs. To simplify labeling, the relative electrical potentials between adenine and thymidine, guanine and cytosine can be represented by graphical labels instead of alphabetical labels. The dsDNA can then be divided into Pyu and Puy dsDNA segments. This study compared Pyu and Puy dsDNAs using different sets of oligo-dsDNAs. The results showed that Pyu oligo-dsDNAs have stronger hybridization potential, a more extended conformation, and increased IR absorption at 3400-3200 $cm^{-1}$ compared to Puy oligo-dsDNAs. Therefore, it is assumed that Pyu dsDNA segments preserve native DNA signals more efficiently than Puy dsDNA segments. All the dsDNA code can be divided into Pyu dsDNA segments, which are expected to preserve the electric charge for different DNA signals.

Before this study, we investigated the magnetic effect on the hydrogen bonding of water, and found characteristic physicochemical properties of magnetically treated water (MTW). The MTW was produced by unidirectional pulsating magnetic field, 800 Gauss, 9 Hertz for 1 hour, and showed the increase of T2 relaxation time in NMR spectroscopy[6], the decrease of electrical conductivity and surface tension in degassed state[6], the increase of IR absorption at 2400-1900 $cm^{-1}$ (peak at 2115 $cm^{-1}$)[6], the increased solubility speed of glycine ($NH_2CH_2COOH$), boric acid ($H_3BO_3$), and $MgSO_4$ but the decreased solubility speed of urea ($CH_4N_2O$), sodium citrate ($HOC(CO_2Na)\text{-}(CH_2CO_2Na)_2\text{-}2H_2O$) and $(NH_4)_2SO_4$[11], the increase of gypsum crystallization[5], the increased precipitation $BaSO_4$, $BaCO_3$, and $CaCO_3$[5], the increase of hydration hardening speed of gypsum plaster[5], the decrease in critical micelle concentrations (CMC) of sodium dodecyl sulfate (SDS), cetyltrimethylammonium bromide (CTAB), and Pluronic F-68[12], the increase of bony decalcification[13], the increase of membrane permeability, and the decrease of free radical activity[12].

The physicochemical properties of MTW may be attributed to the increased strength of hydrogen bonding between water molecules, indicating transient electron transfer from hydrogen donor to acceptor. In addition to the electrostatic interaction, the non-covalent but covalency state of hydrogen bonds may have a memory effect and can increase the dielectric potential in associated molecules[14]. The phenomenon may be explained by the Zeeman effect in weak magnetic fields (<1T). This study applied the magnetic reaction found in the hydrogen bonds of water molecules to the hydrogen bonds for DNA hybridization.

Based on the HBMR between DNA base pairs, we developed the cyclic electromagnetic simulation (CEDS), which was designed to induce electric charge in each A-T and G-C pair by Zeeman effect in weak magnetic field depending on base pair polarities and effective magnetic exposure time for A-T and G-C pairs, 280 msec and 480 msec, respectively. The effectiveness of CEDS may vary depending on the context of dsDNAs in different environments. This study found that Pyu oligo-dsDNAs treated with CEDS using a corresponding sequence showed strong hybridization, unique conformational changes, and increased IR absorption at 3400-3200 $cm^{-1}$ compared to Puy oligo-dsDNAs. It is hypothesized that the Pyu dsDNA segment is a fundamental unit of the entire dsDNA code. The study conducted various experiments to understand CEDS effect on the DNA functions using different Pyu oligo-dsDNAs.

This study compares the hybridization strength, conformation, and IR absorption of Pyu and Puy oligo-dsDNAs using various assays, and found that Pyu oligo-dsDNA exhibits stronger hybridization, a more extended conformation, and increased IR absorption at 3400-3200 $cm^{-1}$ compared to Puy oligo-dsDNAs. These results suggest that Pyu oligo-dsDNAs have a more stable DNA conformation and may contain basic units that play important roles in DNA functions. As a result, different Pyu oligo-dsDNAs were used to evaluate CEDS effects using weak magnetic field, 10-30 Gauss for HBMR in dsDNA in this study.

Since dsDNAs function through binding with other molecules, they maintain amphoteric properties with cationic bases in the center of DNA axis and anionic phosphate groups in the peripheral backbone of dsDNA. Thus, the cationic EtBr and spermidine are representative DNA binding molecules that can alter the conformation of dsDNA. Although they bind to dsDNA differently, they both condense the dsDNA and decrease its conformation. We used various Pyu oligo-dsDNAs in the EtBr and spermidine binding assays with CEDS treatment, and found that the oligo-dsDNAs tended to be expanded by CEDS, resulting in an increase in DNA conformation. The binding assay demonstrated that CEDS using a target sequence can partially increase the conformational size of target dsDNA condensed by EtBr or spermidine, indicating that CEDS can prevent and reverse the condensation of target DNA induced by EtBr and spermidine by increasing the hydrogen bonding strength between base pairs against DNA base stacking.

To facilitate restriction digestion of dsDNA, restriction enzymes (RE) must be able to easily recognize specific RE site sequences in dsDNA. This study found that the specific RE site sequence*-CEDS consistently enhances RE digestion of plasmid DNA. This suggests that CEDS is able to adjust the conformation of the RE site sequence to recruit the RE enzyme. The repeated stimulation of the hydrogen bonds between DNA base pairs by the hydrogen bonding magnetic reaction (HBMR) can lead to an accumulation of electric charge in the associated pyrimidines and purine bases. This can increase the polarity of the base pairs, resulting in a particular DNA conformation that is more likely to be recognized by a specific restriction enzyme.

This study also found that *in vitro* RNA transcription was increased by the specific promoter sequence-CEDS when 20-25 Gauss magnetic field was applied for 30 min to plasmid DNA containing the human VEGF, vWF, or elafin gene compared to the untreated negative control and the positive control using a nonspecific sequence, 12A-CEDS. The promoter sequence-CEDS is expected to activate the target promoter by recruiting more RNA polymerase than the negative and positive controls.

In the *in vivo* experiment using E. coli culture transfected with pE-GFP-1 and pBluescript SK (-) vectors, the promoter sequence-CEDS induced higher levels of GFP and β-galactosidase compared to the negative control and the positive control treated with nonspecific sequence-CEDS. However, the T3 promoter sequence-CEDS increased not only the amount of plasmid DNAs but also the number of surviving cells. The data suggest that the promoter sequence-CEDS can activate the target promoter and enhance RNA and DNA transcription, resulting in the overexpression of corresponding proteins *in vivo*.

In the previous studies, we also examined the effect of pulsed unipolar magnetic fields on cells and animals in relation to bio-hazards of alternating magnetic fields. The results showed that the pulsed unipolar magnetic field, with strength of 63-225 Gauss and a frequency of 120 Hertz, had almost no impact on the culture of human osteogenic sarcoma (HOS) cells until 6 hours after magnetic exposure. After this time, apoptotic cell death was observed[15]. The impact of extremely low frequency electromagnetic field (ELF-EMF) on cells can vary depending on the strength, frequency, and duration of exposure. In a study on mice, exposure to pulsed unipolar ELF-EMF at 730-960 Gauss resulted in rare changes in testicular cells up to 1 hour after exposure, after which the cells gradually underwent apoptotic cell death. Additionally, exposure to 7 Hertz ELF-EMF resulted in more apoptosis of spermatocytes compared to exposure to 1, 20, 40, and 80 Hertz ELF-EMF[16]. After 2 hours of magnetic exposure with a pulsed unipolar magnetic field of 0.2-0.3T and 60 Hertz, there was almost no expression of amyloid precursor protein (APP) in the mouse brain. However, after 4 hours of magnetic exposure, there was an increase in APP expression[17,18].

The results of the *in vitro* and *in vivo* experiments suggest that the damages caused by the pulsed unipolar magnetic field on cells and animals are due to the occurrence of free radicals resulting from the alternating magnetic field. In fact, the pulsed unipolar magnetic field can induce free radicals that can be used for bony decalcification in histological procedures[13]. However, the cells and animals remained healthy and exhibited normal features in gross and histological observations for up to 1-2 hours after magnetic exposure[15-17]. Therefore, it is recommended to limit the strength, frequency, and exposure time of the magnetic field for CEDS in biological applications to prevent potential bio-hazards in cells. In this study, we utilized a pulsed unipolar magnetic field with a frequency of 100 Hertz, strength of less than 30 Gauss, and a duration of less than 30 min, as a standard procedures in CEDS.

This study to develop the HBMR-based gene regulation system can be challenging for many different fields of biomedical sciences. Therefore, it is imperative that the results be carefully verified by further molecular biology and genetic studies by numerous researchers before CEDS effect could be identified as safe and beneficial for humans. This study constitutes Part I of the development of HBMR-based gene regulation, and Part II of the application of HBMR-based gene regulation in the other manuscript.

**Acknowledgments**

We would like to express our gratitude to the late Professor Je Geun Chi and the late Dr. Soo Il Chung, who contributed to this research in part.